\documentclass[useAMS,usenatbib,usegraphicx]{mn2e}
\usepackage{float,color}
\usepackage{epsf,rotating,url}
\usepackage{amssymb,amsfonts,amstext,amsgen,amsopn,amsxtra,indentfirst,lscape,times,rotating}
\usepackage{longtable,subfigure}
\usepackage{graphics}
\usepackage{keyval}
\usepackage{trig}
\usepackage[normalem]{ulem}
\usepackage{dcolumn}
\usepackage{bm}
\def\beq{\begin{equation}}
\def\eeq{\end{equation}}
\def\bey{\begin{eqnarray}}
\def\eey{\end{eqnarray}}
\def\msun{M_\odot}
\def\lsun{L_\odot}
\def\kms{\, {\rm km \, s}^{-1} }

\def\mnras{MNRAS}
\def\apj{ApJ}
\def\nat{Nature}
\def\apjs{ApJ}
\def\apjl{ApJ}
\def\na{New Astronomy}
\def\araa{ARAA}

\def\aap{A \& A}
\def\aj{AJ}

\def\numb{13~}
\def\aap{Astron. Astrophys.}
\def\ser{S\'ersic }
\title{The potential role of NGC 205 in generating Andromeda's vast thin co-rotating plane of satellite galaxies}
\author[Garry W. Angus, Paul Coppin, Gianfranco Gentile, Antonaldo Diaferio]{
\Large Garry W. Angus$^{1}$\thanks{E-mail: angus.gz@gmail.com}, Paul Coppin$^{1}$, Gianfranco Gentile$^{2,1}$, Antonaldo Diaferio$^{3,4}$\\ 
$^{1}$Department of Physics and Astrophysics, Vrije Universiteit Brussel, Pleinlaan 2, 1050 Brussels, Belgium \\
$^{2}$Sterrenkundig Observatorium, Universiteit Gent, Krijgslaan 281, 9000, Gent, Belgium\\
$^{3}$Dipartimento di Fisica, Universit\`a di Torino, Via P. Giuria 1, I-10125, Torino, Italy \\
$^{4}$Istituto Nazionale di Fisica Nucleare, Via P. Giuria 1, I-10125, Torino, Italy\\
}
\begin{document}
\date{\today}
\maketitle
\begin{abstract}
The Andromeda galaxy is observed to have a system of two large dwarf ellipticals and $\sim$13 smaller satellite galaxies that are currently co-rotating in a thin plane, in addition to 2 counter-rotating satellite galaxies. We explored the consistency of those observations with a scenario where the majority of the co-rotating satellite galaxies originated from a subhalo group, where NGC 205 was the host and the satellite galaxies occupied dark matter sub-subhalos. We ran N-body simulations of a close encounter between NGC 205 and M31. In the simulations, NGC 205 was surrounded by massless particles to statistically sample the distribution of the sub-subhalos expected in a subhalo that has a mass similar to NGC 205. We made Monte Carlo samplings and found that, using a set of reference parameters, the probability of producing a thinner distribution of sub-subhalos than the observed NGC 205 + 15 smaller satellites (thus including the 2 counter-rotators, but excluding M32) increased from $<10^{-8}$ for the initial distribution to $\sim10^{-2}$ at pericentre. The probability of the simulated sub-subhalos occupying the locations of the observed co-rotating satellites in the line of sight velocity versus projected on-sky distance plane is at most $2\times10^{-3}$ for 11 out of 13 satellites. Increasing the mass of M31 and the extent of the initial distribution of sub-subhalos gives a maximum probability of $4\times10^{-3}$ for all 13 co-rotating satellites, but the probability of producing the thinness would drop to $\sim 10^{-3}$.

\end{abstract}

\section{Introduction}
\protect\label{sec:intr}
In the concordance cosmological model (\citealt{davis85,frenk85,frenk88,spergel07,planck14}) large spiral galaxies, like the Milky Way (MW) and Andromeda (M31), form in an expanding universe through the aggregation of cold dark matter around an initial seed overdensity (\citealt{white78,white91}). The smallest scale perturbations attract the surrounding matter in the shortest time-scale and thus form their halos first. Owing to the higher background density at earlier times, these small halos are more dense than their larger counterparts. They are eventually encompassed by larger halos and some are destroyed by tidal forces causing their dark matter to be dispersed throughout the smooth component of the larger halo. A portion survive the tidal forces and can be labelled subhalos.

The Aquarius project (\citealt{springel08}) is one of the largest cosmological N-body simulations of Milky Way/Andromeda sized dark matter halos, with a thorough investigation of subhalo statistics. According to their simulations, these subhalos have a more extended distribution around the host halo than the smooth component. This is inevitable since the subhalos at great distances from the host halo experience weaker tides and are therefore less prone to the culling that the subhalos on short radius orbits encounter. The mass function of subhalos surrounding the host halo can on average be represented by a power law $dn/dm \propto m^{-1.9}$ and typically the most massive subhalos are around one hundredth the mass of the host. Those most massive subhalos also tend to lie beyond the virial radius of the host.

These massive subhalos can also play host to sub-subhalos (see \citealt{diemand07}), which are halos that are bound primarily to the subhalo. The mass function of these sub-subhalos is similar to that of the subhalos, albeit with a slightly lower normalisation, meaning fewer total satellites per decade of mass. The distribution of the sub-subhalos has not been precisely quantified and although it may be similar to that of the subhalos, its extent is likely truncated given the tidal forces of the host halo (which can sunder the sub-subhalo from the subhalo). In addition, sub-subhalos cannot be replenished.

It is well known that only a fraction of the possible subhalos of the MW and M31 host satellite galaxies (\citealt{klypin99,moore99}). Furthermore, some of the most massive predicted subhalos of those two large galaxies do not have corresponding stellar (luminous) counterparts. This means the bright satellite galaxies of the MW and M31 are hosted by subhalos that are much less massive than the most massive ones. This contradiction is known as the too big to fail (TBTF) problem (\citealt{boylan11,boylan12,garrison13,garrison14}). In essence, if a low mass subhalo is able to accrete gas and form stars with its puny potential well, then so should a more massive subhalo.

A further complication is that roughly half of the detected satellite galaxies orbiting M31 (see e.g. \citealt{koch06,mcconnachie12,conn13}) lie in a highly flattened plane and the majority of those (14 out of 16) have a common sense of rotation (\citealt{ibata13}) - dubbed the vast thin co-rotating (hereafter VTC) plane of satellites. This highly organised distribution is unlikely to be the result of stochastic star formation in certain subhalos (\citealt{walker14}  and \citealt{ibata14a} in response to \citealt{bahl14}). As it happens, the MW has a similar satellite distribution where most satellite galaxies could be part of a highly extended and flattened disk - some 300~kpc in radius with a root mean squared (RMS) thickness of merely 30~kpc (\citealt{lb83,metz08,kroupa10,pawlowski15,pawlowski16}). Unfortunately, the difficulties involved with measuring proper motions make it challenging to determine if the satellites are co-rotating.

A matter of current debate in the literature is whether co-rotating planes of satellites are found around other large galaxies (\citealt{ibata14,cautun15a,cautun15b,phillips15,ibata15}). This is important to confirm since it will determine whether this is a generic property of satellite distributions, or if the Local Group is simply peculiar. \cite{pawlowski13} have shown that there is evidence of a link between planes of galaxies and satellites within the local group.

Proposed theories to explain this contrived distribution include mergers between M31 and other galaxies (\citealt{sawa05}), which result in tidal dwarf galaxies that would form in a plane (\citealt{kroupa97,gentile07a,hammer13,hammer15}). However, this appears unlikely in a cold dark matter framework due to the implied lack of cold dark matter in the tidal dwarfs, which strongly contradicts inferred dark matter abundances of the observed dwarf spheroidals (\citealt{mateo98,walker07,strigari08,strigari10}).

Another suggestion has been that the satellite galaxies were accreted along filaments that feed the host galaxy (\citealt{zentner05,libeskind05,lovell11,wang13,tempel15,libeskind15}). Although all satellites would have coherent motions, the probability of producing the highly ordered system of satellites found in the Local Group has been argued against by \cite{pawlowski12,pawlowski14}. It also offers no explanation for the apparently short duration these satellites have been accreted over. Other studies have suggested it would be difficult to produce the satellite galaxies with low orbital angular momentum in this fashion (see e.g. \citealt{adk11}). 

Another prospect is for a group of galaxies to be accreted (\citealt{donghia08,lihelmi08}). This would have the interesting property of ensuring the satellite galaxy distribution was relatively compact and has the potential to produce low angular momentum galaxies during the encounter between the host and the group. This idea was argued against by \cite{metz09} on the basis that in the study of \cite{tully06} there are no observed galaxy associations that are as thin as the MW's disk of satellites. Although it is true that the RMS projected radii of the galaxy associations identified by \cite{tully06} ($\sim$~150~kpc) are far larger than the thickness of the MW's disk of satellites, that does not rule out the possibility that a component of a galaxy group is more compact. In fact, the majority of the brightest galaxies in each group have a very nearby companion of much lower luminosity, as we demonstrate later. Moreover, the pertinent observation is whether the tight groups proposed by \cite{donghia08} exist at moderate redshifts. It is conceivable that a high fraction of all tight groups have been tidally destroyed by redshift zero. This group infall scenario has regained popularity recently with several studies pointing out the possibility that a reasonable fraction of the Local Group's satellites originally formed in groups (\citealt{deason15,wetzel15,wheeler15,yozin15,smith15}).

A further peculiarity with regards to the satellite system of Andromeda is that the satellites that are part of the co-rotating thin plane are consistent with lying only in an arch surrounding M31, they very possibly do not encircle it. Thus, it makes sense to investigate scenarios where the satellites would intrinsically lie in such a distribution. In any case where the satellites fall from a large distance, they will execute a parabolic orbit. Depending on the impact parameter and the mass of Andromeda, this could result in satellites observed mostly in an arch around M31. This scenario has to satisfy the 3D positions of the satellites relative to M31 and their line of sight velocity.

Here we investigate the likelihood of a variant on the group infall scenario such that, instead of a large group centred on an M33 or LMC size host (\citealt{deason15,wetzel15,sales16}), a subhalo of Andromeda along with all of its sub-subhalos form significant quantities of stars and they execute a fly-by orbit that brings them close to M31. During the close encounter, the satellites are strewn out around M31. Using N-body simulations, we explore the probability that these sub-subhalos now represent the majority of the co-rotating satellites of Andromeda that lie in a thin arch. An idea somewhat similar to this one was simultaneously considered by \cite{smith15} who concentrated on a 1:2 mass ratio encounter between two MW/M31 mass galaxies and the survivability of such a co-rotating plane of satellites, but did not concentrate on the likelihood of their scenario reproducing the observed positions and velocities of the observed satellites. Here we explicitly focus on a statistical analysis of our scenario matching the detailed observed properties of M31's VTC plane of satellites.

In \S\ref{sec:sub} we introduce the subhalo scenario and the simulation setup, in \S\ref{sec:compar} we discuss the satellite sample and their re-orientation to facilitate comparison with the simulations. In \S\ref{sec:res} we present our statistical comparison of the sub-subhalo distribution with the observed satellite distribution of Andromeda and in \S\ref{sec:conc} we discuss the results and draw our conclusions.

\section{The subhalo scenario}
\protect\label{sec:sub}
Our ansatz is that due to photo-ionisation effects (see e.g. \citealt{efstathiou92,bullock00,benson02,dijkstra04,slater13,wetzel13,boylan14}), only the latest accreted subhalos had substantial numbers of bright sub-subhalo companions. Any subhalos accreted earlier would have suffered from the bright UV halo originating from M31, stifling the formation of stars in their sub-subhalos. This would require a sort of patchy re-ionisation, perhaps the kind suggested by \cite{castellano16} or \cite{sharma16}. Any subhalo that weighed more than a few percent of the host would similarly quench their own sub-subhalos, and any subhalo weighing much less than 1\% of the host would not have a significant, observable sub-subhalo distribution. In principle, this could leave a reasonably small window of subhalo masses with accompanying, bright sub-subhalos.

The advantage of an accreted subhalo over a generic group is that the sub-subhalo mass function might closely reflect that of the observed satellites in the VTC plane (see \S\ref{sec:lumfunc}). Furthermore, the mass of the subhalo might reflect the mass of a larger satellite at the centre of the observed satellite distribution. A corollary of this is that a significant fraction of galactic satellites could originate from subhalo groups on long period orbits.

In our scenario, NGC~205 (M110) was the brightest galaxy of one of M31's largest subhalos. Thus, the stellar component of NGC~205 sat at the centre of the subhalo. This subhalo was surrounded by hundreds of sub-subhalos, of which only a fraction (the most massive) produced large quantities of stars. It is our assertion that after a close pericentric passage of M31, the sub-subhalos were spread out along the orbital direction, but tidally compressed in the direction transverse to the orbit. If this effect is strong enough, under certain conditions, then this could in principle explain the VTC plane of satellites surrounding M31 (\citealt{ibata13}).

\subsection{The simulation coordinate system}
\begin{figure*}
\includegraphics[width=17.0cm]{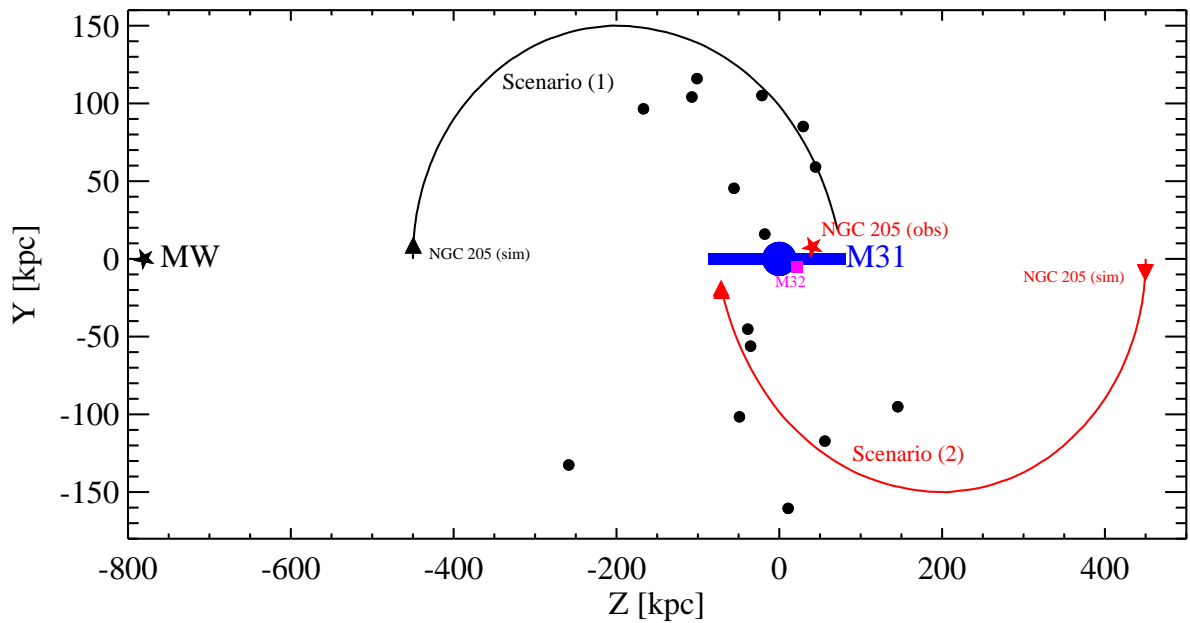}
\caption{The positions of the large galaxies (the Milky Way, M31, M32 and NGC 205) and observed satellites (black circles) in the plane of the proposed orbit of NGC 205 around M31. We show the orbital paths of the two scenarios (1 with a black line and 2 with a red line). The $Z$ coordinates for NGC 205 and M32 are the maximum likelihood values from McConnachie et al. (2005), but their errors may be as high as $\sim70$~kpc (see \S\ref{sec:d205} and de Grijs \& Bono 2014).}
\label{fig:orbitYZ}
\end{figure*}

\protect\label{sec:scs}
To visualise the geometry of the scenario, we have plotted two orthogonal views. The first, shown in Fig~\ref{fig:orbitYZ}, is the $Y$-axis versus the $Z$-axis, which is the orbital plane that NGC~205 moves in. The $Z$-axis is set by the line between the MW (Y, Z)=(0~kpc,-783~kpc) and M31 (Y, Z)=(0~kpc,0~kpc). The $Y$-axis is simply the height above or below M31. We test two different scenarios. The first, which we call scenario 1, is that NGC 205 began several Gyr ago from a position between M31 and the MW. It initially moved in the positive $Y$ direction and finally ended now just behind M31 moving in the negative $Y$ direction. Alternatively, we could have scenario 2 where NGC 205 initially sits several hundred kpc behind M31 and orbits in the negative $Y$ direction. It orbits towards M31 and the MW and currently sits before M31, on the MW's side, and is moving in the positive $Y$ direction. The difference between the two scenarios is that (1) requires NGC 205 to sit just beyond M31 and (2) just in front of it. 

\subsubsection{Distance to NGC 205}
\protect\label{sec:d205}
One might suspect that the uncertainties on the distance to NGC 205 could easily allow both possibilities, however, according to \cite{mcconnachie05} the distance to NGC 205 using the tip of the red giant branch (TRGB) method is $d_{NGC205}=824\pm27$~kpc. Using a distance to M31 of  $d_{M31}=783$~kpc this gives a separation of around 41~kpc and thus a 1$\sigma$ range of roughly 14 to 68~kpc (meaning NGC 205 is behind M31). Alternatively, \cite{degrijs14} give recommended values in their review of distances to Andromeda galaxies with larger errors. From their table 4, the 1$\sigma$ range for the separation between M31 and NGC 205 using the TRGB is given as -67 to 75~kpc, thus allowing NGC 205 to be in front of or behind M31. We thus consider both scenarios.

The majority of the satellite galaxies with $Y$ coordinate greater than 0 (or above M31 and NGC 205) are moving with a line of sight velocity away from the MW and those with a negative $Y$ coordinate (below M31 and NGC 205) are mostly moving towards the MW. 

\begin{figure}
\includegraphics[width=8.50cm]{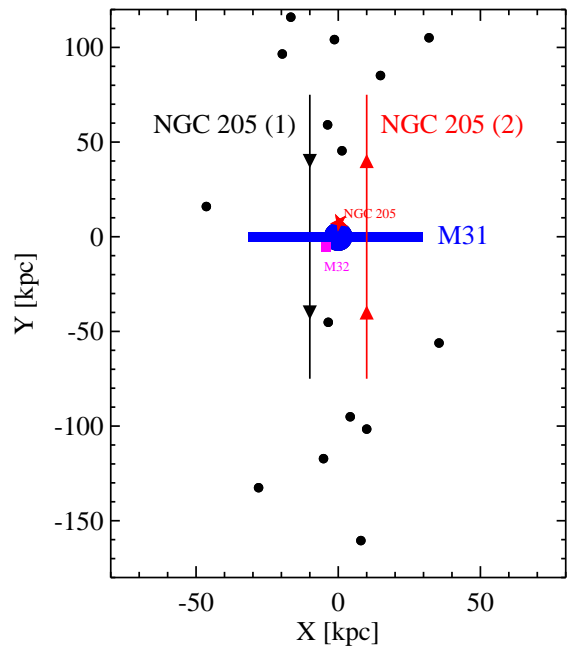}
\caption{The positions of the large galaxies (M31, M32 and NGC 205) and observed satellites (black circles) in the plane of the sky. The Milky Way is located at $Z=-783$~kpc (out of the screen). We show the orbital paths of the two scenarios near pericentre (1 with a black line and 2 with a red line).}
\label{fig:orbitYX}
\end{figure}

The second view, Fig~\ref{fig:orbitYX}, is from the MW towards M31, showing us the plane of the sky. Here we see the recent orbit of NGC 205 through the thin plane of satellites which has a thickness which is surprisingly small in the $X$ direction.

We consider NGC~205 and not M32 because the former's lower velocity relative to M31 makes it a better candidate. Recent observations and theoretical models of M32 suggest it is actually on the Milky Way's side of M31 (\citealt{jensen03,karachentsev04,dierickx14}), but again this distance estimate is likely as uncertain as that of NGC 205.

\subsection{Mass models for M31 and NGC 205}
To simulate this scenario we make N-body realisations for M31, NGC~205 and its dark matter subhalos. M31 and NGC~205 are modelled with four components each: exponential stellar and gaseous disks, a \ser bulge (\citealt{graham05}) and an NFW dark matter halo (\citealt{nfw97}). 

We preface this by saying, besides the fairly well constrained dark halo mass of M31, the only particularly important parameters are the extent of NGC 205's distribution of sub-subhalos, and its dark halo mass. The baryonic parameters and the details of the dark matter distribution have very little influence on the outcome, but are included to add more realism to the simulations. In table~\ref{tab:sbpar} and the following discussion we describe the reference parameters for the mass models. It is to these values that variations, and references, are made in the remainder of the article. 
\subsubsection{M31}
The M31 parameters were chosen to reflect the measured rotation curve from \cite{kent89,widrow05} (but see also \citealt{carignan06,corbelli10,tamm12}). We use slightly tuned versions of the parameters used by \cite{dierickx14}, in particular a \ser bulge (as opposed to a Hernquist bulge; \citealt{hernquist90,baes02}) with a larger mass to better fit the inner rotation curve. The total dark halo mass of $M_{\rm 200}=1.78\times10^{12}\msun$ exceeds that found by the other analyses. One reason for this is that we do not consider globular clusters or satellites good tracers of the galactic potential, for reasons evident from this article.

\subsubsection{NGC 205}
The NGC 205 parameters had relatively more freedom since its dynamics are less precisely constrained.  Since it has likely suffered from the tidal influence of M31 during the orbit, matching the detailed dynamics using modelling like \cite{derijcke06} or \cite{geha06} seemed beyond the scope of our paper. The parameters for NGC 205's dark matter halo are simply chosen to give it $\sim$1\% the mass of M31's dark matter halo, i.e $1.8\times10^{10}\msun$, to ensure that the stellar component of NGC 205 occupies one of M31's largest subhalos (\citealt{springel08}). Of course there is no strict requirement for NGC 205 to occupy the most massive subhalo, so this remains a parameter with a substantial amount of uncertainty.

The radial distribution of the dark matter halo is chosen to give it a similar extent to the most massive subhalos of the Aquarius project, thus it is similar to the dark halo profile of M32 from \cite{dierickx14} - which is a comparable dwarf galaxy in terms of its luminosity. The two most massive subhalos surrounding the $M_{\rm 200}\sim1.8\times10^{12}\msun$ host halo (the same as our dark matter halo mass for M31)  in \cite{springel08} have $r_{250}=56.7$ and 55.3 kpc, whereas \cite{dierickx14} chose $r_{200}=67$~kpc. We use $r_{200}=60$~kpc, but again, the extent of the dark matter halo is less relevant than its overall mass since the sub-subhalos orbit at large distances.

For the baryonic components of NGC 205, nominal values were selected for broad agreement with the luminosity and size.  According to \cite{mcconnachie12}, the V-band magnitude for NGC 205 is -16.5, which corresponds to a luminosity of $\sim 3\times 10^{8}\lsun$. These luminosities can often be under-estimated due to insufficiently modelled extinction or the difficult to detect outer surface brightness profile (\citealt{crno}), but we assume that these uncertainties are included in the unknown stellar mass-to-light ratio - which we set to $M/L_V\sim2.7$ (e.g. \citealt{belldejong}). Thus the dark matter to baryonic (stars plus gas) mass ratio is 18.2, which is a reasonable value for such a galaxy and is just slightly higher than M31's value of 16.8.

The stellar mass of NGC 205 is split almost evenly between bulge and disk (as \citealt{dierickx14} did for M32), even though it is a dwarf elliptical galaxy, since the galaxy might have changed morphology during the orbit.
The exponential scale-heights of the stellar and gaseous disks are found from the scaling relation $h_z \approx 0.2 h_R^{0.633}$ (see \citealt{bershady10b,angus15}).
\subsubsection{Velocity distributions}
The velocity distributions for all spherical components are sampled from their distribution functions. The distribution functions are found using potentials that assume a spherical gravitational field for the disks. The NFW halos are generated using the truncation devised by \cite{kazantzidis04} where we set $r_{\rm dec}=r_{\rm 200}$. The disks themselves were initialised using the method of \cite{hernquist93}. The only difference is that an exponential vertical density distribution is assumed here in place of a $sech^2$ distribution.

\subsubsection{Sub-subhalo distribution}
\protect\label{sec:eps}
The final component is the distribution of sub-subhalos surrounding NGC 205. Note that we do not assume that the sub-subhalos trace the dark matter distribution. Instead, we start with the Einasto number density distribution of subhalos around their host found by the Aquarius project (\citealt{springel08}), which goes like 
\beq
\protect\label{eqn:nsub}
n_{\rm sub}\propto \exp\left(-{2 \over \alpha}\left[\left({r\over r_{-2}}\right)^{\alpha}-1\right]\right).
\eeq
\cite{springel08} do not report the distributions of sub-subhalos, presumably because they are quite stochastic. However, they make it clear that the distribution of sub-subhalos around their subhalo host is not expected to be exactly a scaled down version of Eq~\ref{eqn:nsub} because of tidal stripping and the fact the sub-subhalos cannot easily be restocked, however, it is the most reasonable starting point. 
We previously set the virial radius of NGC 205 to $r_{\rm 200}=60$~kpc. Per \S3.2 of \cite{springel08}, we define the sub-subhalo distribution by setting $\alpha=0.68$ and $r_{-2}=0.81 r_{\rm 200}$. 

The influence of the parameter $r_{-2}$, as well as the masses of both galaxies' dark matter halos and the orbital pericentre are investigated in the results section (see \S\ref{sec:res}).

The sub-subhalo distribution extends well beyond the virial radius of NGC 205 (as far as 400~kpc). Although the actual number of massive sub-subhalos (say $M_{\rm sub-sub}/M_{\rm sub}>10^{-5}$) should only be of the order of 100, we simulate 200k in order to statistically compare with the observed satellites. Again, the velocity distribution is sampled from its distribution function, but the particles (each representing a sub-subhalo) are effectively massless. 

\subsection{Stability and simulation setup}
To demonstrate the stability of the various components, we ran simulations of M31 in isolation and NGC 205 (with the orbiting sub-subhalos) in isolation for 10~Gyr. Each galaxy was resolved with $10^6$ particles and the particle masses were identical for each mass component. All simulations were carried out using the Gadget-2 N-body code (\citealt{springel05}). Plots showing the limited evolution of each mass component are given in the appendix (Figs~\ref{fig:evo1}-\ref{fig:evo3}).  Apart from some thickening of the disk components, the galaxies appear very stable. Of most relevance, the dark matter halos are highly stable, as are the sub-subhalos.

To test the hypothesis that the VTC plane of co-rotating satellites originates from a close tidal encounter between M31 and NGC 205, we need a simulation where such a close encounter occurs. To facilitate this, we give NGC 205 an initial offset of 450~kpc from M31 and a small tangential velocity. This means NGC 205 is initially at apocentre and will orbit towards a pericentre much closer to M31, as per Fig~\ref{fig:orbitYZ}. A distance of 450~kpc was chosen to be sufficiently far away to give the tidal forces enough time to work on the sub-subhalo distribution, assuming that any further away the tidal forces would be minimal. 

The range of initial tangential velocities we probe are between 25 and 80~$\kms$. In \cite{diemand04} they show (with the red circles in the bottom right panel of their figure 3) the 3D velocity dispersion of subhalos as a function of radius normalised by the peak circular velocity of the dark matter halo (which here is around 200~$\kms$). Extrapolating this velocity dispersion to 2 virial radii should give the 3D velocity dispersion of a subhalo. Since the tangential motion is two out of three of these dimensions, it is reasonable to use the 3D velocity dispersion as an approximation. Thus the expected tangential velocity dispersion for our initial conditions would be somewhere around $130\pm30~\kms$. Assuming a Gaussian distribution for the tangential velocities with the aforementioned tangential velocity dispersion and comparing with the cumulative distribution function, the fraction of orbits within the range of velocities we probe is thus around 0.31 and within a more conservative velocity range of 30-65~$\kms$ the fraction is around 0.2. Thus the velocity ranges we consider are fairly typical of the expected velocities of subhalos that are 2 virial radii from their host.

 We also tested larger initial offsets and found no meaningful changes in the outcome. Ideally, this simulation should take place in a cosmological setting, but we assume here that the difference in orbital path and impact velocity are minor.

What is of relevance is not the initial offset, although this must be large enough to produce a suitable pericentric velocity, but rather the pericentre, or impact parameter.  In order to affect different pericentric distances, or impact parameters, we simply vary the initial tangential velocity of NGC 205. Throughout the full orbit, but with greater cadence once NGC 205 is near pericentre, we compare the simulated positions and velocities of the sub-subhalos with the observed satellites of M31.

\section{Comparing simulated sub-subhalos with observed satellites}
\protect\label{sec:compar}

\subsection{Which satellite galaxies to include in the statistical sample}
For the statistical comparison between the simulated sub-subhalos and the satellites, one of the key questions is which satellite galaxies to include.

The satellite galaxies that \cite{ibata13} showed to be located in a vast, thin (but not exclusively co-rotating) plane around M31 are M32, NGC 147, 185, 205 and AND I, III, IX, XI, XII, XIII, XIV, XVI, XVII, XXV, XXVI, XXVII \& XXX (also known as C2). AND XXXII \& LGS 3, which are catalogued by \cite{mcconnachie12}, are not included since they lie beyond the Pan Andromeda (PANDA) footprint where the vicinity is not fully surveyed. We do not include M32 since it is too massive to be part of the NGC 205 group. It is thus a shortcoming of our scenario and M32 would have to be an interloper. Alternatively, it could be bound to NGC 205 and co-orbiting with it like the two Magellanic Clouds. Or, it could be in a short period orbit of M31 given its small offset and line of sight velocity.

NGC 205 is obviously not included because it is a prior of the statistical calculation that it be close to the origin in the VTC plane. Four other satellites are handled on a case by case basis since, mainly for the line of sight velocity, they are less consistent than the others and we wish to gauge how significant their inclusion or exclusion is. AND XIII and AND XXVII do not share the same sense of rotation as the other satellites and are thus highly unlikely to be prior members of the subhalo group. Similarly, although AND XII \& XXVI orbit with the correct orbital sense, their velocities appear somewhat less consistent with the scenario than the remaining 11.

On the one hand, these counter rotating satellites mean that our explanation for the thin disk of co-rotating satellites is incomplete. The alignment of AND XIII \& XXVII with the plane of satellites would have to be coincidental. On the other, this would have to be the case for the majority of models attempting to explain the origin of the VTC plane of satellites, even modified gravity models creating tidal dwarf galaxies. Nevertheless, we still include them in our analysis of the likelihood of having so many satellites in a thin plane since even if they do not originate from a subhalo group centred on NGC 205, their coincidental alignment with the plane must be accounted for statistically.

Therefore, we have 13 satellite galaxies that are part of the VTC plane (including NGC 147, NGC 185, AND XII, and AND XXVI); another 2 (AND XIII and AND XXVII) that are in the plane, but not co-rotating; M31 and NGC 205, which are expected to be there as the prior of the scenario; and M32 which is assumed to be in a short period orbit of M31 and therefore neglected.

\subsection{Subhalo group luminosity function}
\protect\label{sec:lumfunc}

\begin{figure}
\includegraphics[width=8.5cm]{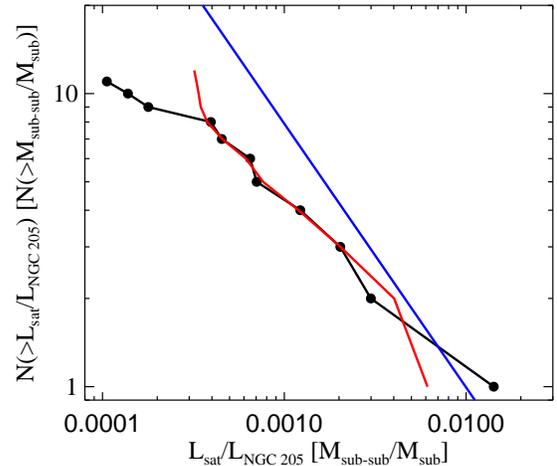}
\caption{The filled black circles and connecting black line show the observed luminosity function of the satellite galaxies we propose originate from a subhalo group and now comprise the majority of the VTC plane of satellites. The red line is a simulated mass function of sub-subhalos from a specifically selected subhalo group (Springel et al. 2008). The blue line is the typical mass function of subhalos around a host.}
\label{fig:lumfunc}
\end{figure}

The advantage of using NGC 205 as the subhalo host of a distribution of sub-subhalos is that we have a reasonable expectation of the mass function of sub-subhalos. In Fig~19 of \cite{springel08}, they show the cumulative mass function of sub-subhalos within several subhalos with each panel displaying a different subhalo.
In Fig~\ref{fig:lumfunc} we show the observed cumulative luminosity function for the 11 satellite galaxies we assume comprised the subhalo group that became the VTC plane of satellites. This group excludes NGC 147 and 185 since they are too massive to be sub-subhalos bound to NGC 205, as well as AND XIII \& XXVII since they are counter rotating. The former two satellites are often considered to be tidally bound to each other (however, see \citealt{evslin14}), but are far too bright to have formed in the subhalo of NGC 205, i.e. to be associated with a sub-subhalo. The most logical explanation would be that they are each associated with another, less massive subhalo of M31 and became bound to NGC 205 during their orbit towards M31, similar to the LMC and the SMC and NGC~147 \& 185 themselves.

The luminosities for each galaxy are taken from the updated list of \cite{mcconnachie12}. Over the luminosity function of the observed satellite galaxies we plot the simulated mass function of sub-subhalos for a specifically chosen subhalo from \cite{springel08}, assuming that the ratio between satellite and host luminosity equals their dark matter halo mass ratios. This is the second subhalo group from their Fig~19 and it is given a red line as per the original plot. To a large extent this simulated mass function resembles the observed luminosity function, thus we can say there is some evidence to suggest the luminosity function of the observed satellite galaxies is representative of the expected mass function of sub-subhalos from simulations. We also overplot the standard mass function of subhalos around a host galaxy (blue line), which can of course be determined with far greater precision.

\subsection{Re-orienting the observed satellite galaxies for comparison}
\begin{figure}
\includegraphics[width=8.5cm]{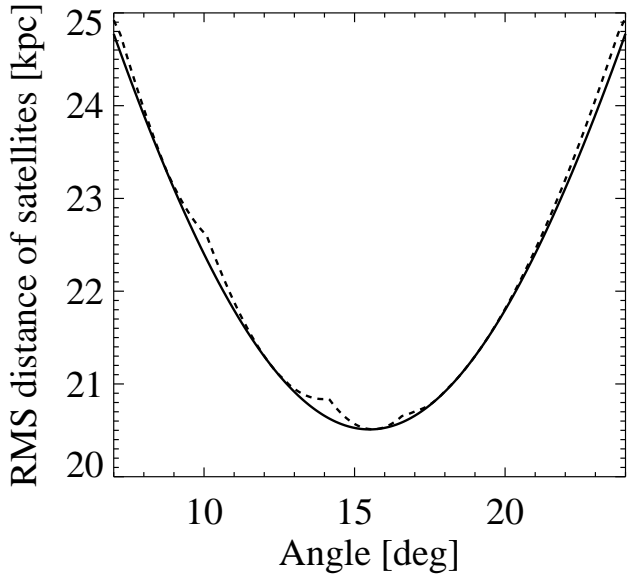}
\caption{The RMS distance of the 15 observed satellites in the VTC plane is plotted against rotation angle. The RMS distances are computed relative to the mean $X$ position of the observed satellites (solid line) and the median (dashed line).}
\label{fig:rmsa}
\end{figure}

The observed distribution of satellite galaxies that belong to the co-rotating plane of satellites, as shown in Fig~1 of \cite{ibata13}, is not exactly aligned with the vertical direction on the graph. Our simulations are setup such that NGC 205 and M31 are initially separated along the $Z$ direction and movement is exclusively in the $Y$-$Z$ plane (see our Figs~\ref{fig:orbitYZ}-\ref{fig:orbitYX}). Therefore, NGC 205 has no motion in the $X$ direction and our key comparison is between the $X$ distribution of the satellites and sub-subhalos. To make comparison with the simulations easier, we simply rotate the observed satellites around the origin in the $X$-$Y$ plane, which is chosen to be M31. In Fig~\ref{fig:rmsa} we plot the RMS distance of the 15 observed satellites from the $Y$-axis as a function of rotation angle. We do this for two cases, one where the RMS $X$ distance is calculated relative to the mean $X$ position of the satellites and another relative to the median. The rotation angle that minimises the RMS distance from the median $X$ position of the observed satellites is 15.6$^{\circ}$ and we rotate all satellites in the $X$-$Y$ plane by this angle and subtract the median $X$ value for all satellites.

To further align the observed satellites with the $Z$-$Y$ plane of the simulations, we subtract 1.26~kpc from their $X$-coordinates. We do this to place the origin close to the centre of the VTC plane of satellites, but importantly midway between the two satellites with the smallest distance from the plane (i.e. smallest $|X|$ values). This shift is important because if we place one satellite at $X=0$ it becomes meaningless to compute how many simulated sub-subhalos have a smaller $|X|$-coordinate value.

This means that the observed NGC 205 sits at ($X_{\rm NGC 205}=0.25~kpc$, $Y_{\rm NGC 205}=7.7$~kpc) and half the satellites have larger $Y$ values and half have lower. The observed M31 is at ($X_{\rm M31}=-6$~kpc, $Y_{\rm M31}=0$~kpc) and the $X$ coordinate is the only one (including $Y$, $Z$ and $V_Z$) for which M31 is not central. In the simulations we ignore this subtlety and set $X=0$~kpc for M31.

The extra constraint we have is that NGC 205 is currently moving with line of sight velocity relative to M31 of $V_Z=54~\kms$ and therefore the ideal instant to make any statistical comparison is when the simulated NGC 205 has its observed $Y$ and $V_Z$ values. Of course this only happens once per orbit. Later when we refer to making comparisons near pericentre, we mean at this precise moment. When we wish to make a statistical comparison along the orbit, we will ignore the line of sight velocity whilst ensuring the height of the simulated NGC 205 is very close to the observed 7.7~kpc.

\subsection{Re-orienting the simulated sub-subhalos for comparison and which regions to exclude}
\protect\label{sec:mask}

To compare the simulated sub-subhalos with the observed satellites, we set the simulated M31 as the origin and we rotate NGC 205 and all its sub-subhalos in the $Z$-$Y$ plane until the $Y$ coordinates of the observed NGC 205, M31 and the Milky Way (7.7~kpc, 0~kpc and 0~kpc respectively) are matched by the simulated ones. This allows us to compare the distribution of sub-subhalos at any time, but the observed fact that NGC 205 is nearly in-line with M31 along the $Z$ direction is fixed. It also ensures that close to half of the sub-subhalos are above NGC 205 (larger $Y$ values) and half are below, as is observed.

Calculating any statistics from the distribution of sub-subhalos requires some pre-filtering because the majority of the sub-subhalos with low $|X|$ values have low $Y$ values, i.e. the distribution is narrowest close to M31. However, no satellites are observed next to M31, in fact the closest satellites to M31 are beyond 40~kpc. Therefore, we assume that there are satellites close to M31, but they are just not detected. In order not to bias our statistics by considering sub-subhalos in the region close to M31 where no satellites are found, we exclude all sub-subhalos with $\sqrt{X^2+Y^2}<40$~kpc. We also exclude all sub-subhalos with $Y>135$~kpc and $Y<-180$~kpc, which are roughly the limits of the PANDAS footprint. These exclusion criteria are applied to all analyses, even if they do not directly concern the $X$ coordinate and we discuss their influence in \S\ref{sec:ydir}.


\subsection{Reference simulation}
In table~\ref{tab:sbpar}, we list the reference set of parameters for our simulations. We refer to these parameters when listing the parameters of all other simulations. In the next section we use scenario (2) from Figs~\ref{fig:orbitYZ}-\ref{fig:orbitYX} with an initial tangential velocity for NGC 205 of $V_Y$=-60~$\kms$, which in this case means a pericentre of $Z=-72$~kpc, to display some basic results and properties of the simulations.

For this reference simulation, we plot in Fig~\ref{fig:xya} the on-sky projection of the observed M31 satellites, M31 and NGC 205. The simulated sub-subhalos are plotted at the point near to pericentre when the simulation phase space coordinates of NGC 205 agree with the observed ones. The location of the observed M31 is shown by the blue galaxy, and only differs from the simulated position for the $X$ coordinate: in simulations $X_{\rm M31}=0~\rm kpc$, whereas the observed M31 relative to the other satellites is $X_{\rm M31}=-6$~kpc. The simulated NGC~205 is identified by the red star and the only difference between simulated and observed coordinates is that the true line of sight distance $Z_{NGC205}$ is more open to debate than the other coordinates (see \S\ref{sec:d205}). 

This pericentric distribution of sub-subhalos in Fig~\ref{fig:xya} can be compared with the distribution at the beginning of the simulation before any tidal distortion (Fig~\ref{fig:xyb}). Clearly the close pericentric passage significantly narrows the sub-subhalo distribution in the $X$-direction and greatly extends it radially in the $Y$-direction.

\section{Results}
\protect\label{sec:res}
In what follows, we discuss a series of statistical tests chosen to estimate the likelihood that the sub-subhalo scenario led to a thin plane of satellites, as well as the observations of the other phase space coordinates ($Y$, $Z$ and $V_Z$). 
\subsection{Tests of the X distribution / plane thickness}

\subsubsection{Kolmogorov-Smirnov test}
\protect\label{sec:ks}

\begin{figure}
\includegraphics[width=8.5cm]{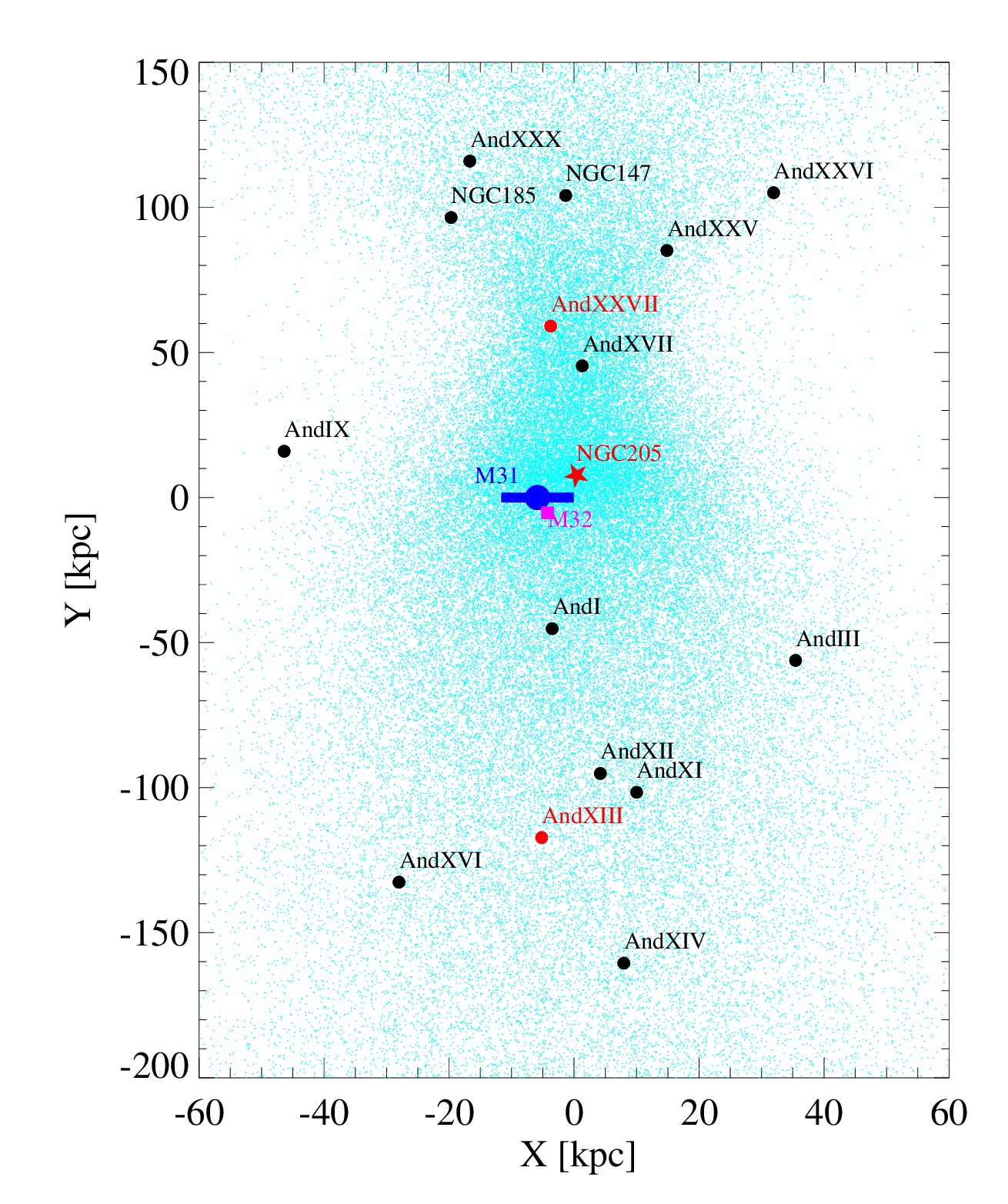}
\caption{The sub-subhalos and observed satellites in the $X$-$Y$ plane (i.e. perpendicular distance from the VTC plane and plane of the sky location above or below M31) at the time when the observed and simulated coordinates for NGC 205 are identical. The 13 observed satellites that we propose originated from a subhalo group are represented by the black filled circles and their names are indicated. The turquoise dots signify the simulated sub-subhalos. The red star is the location of the observed NGC~205 (identical to the simulated location). Although we plot the observed location of M31 with the blue galaxy shape, the simulated M31 is at the origin. M32 is located at the pink square, while AND XIII \& XXVII are shown as red circles because they are not considered to have originated from the subhalo group. }
\label{fig:xya}
\end{figure}

\begin{figure}
\includegraphics[width=8.5cm]{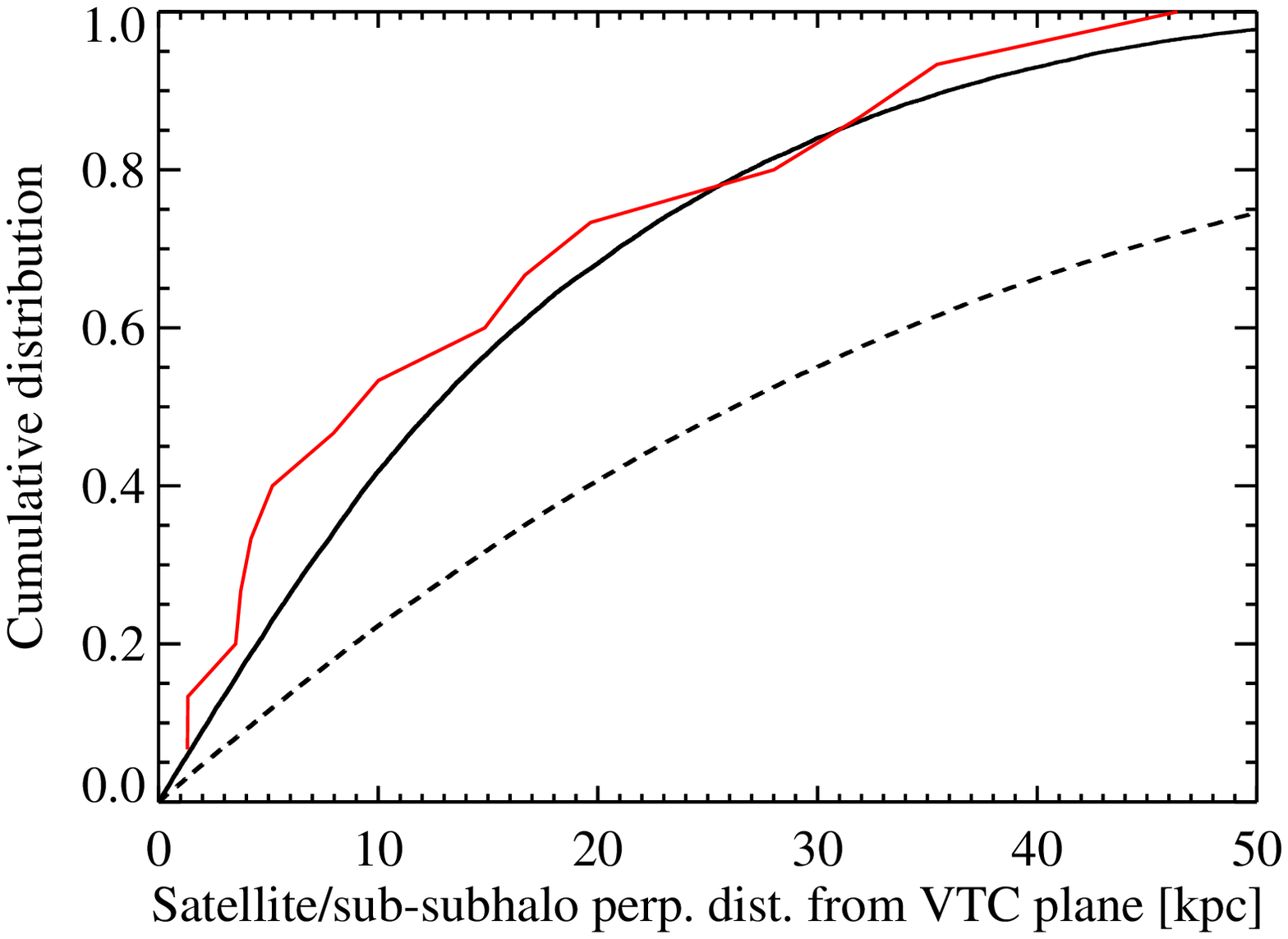}
\caption{Cumulative distribution of distances, $|X|$, perpendicular to the VTC plane for our reference simulation. The solid red line includes all 15 observed satellites, including the counter rotating AND~XIII \& XXVII, but excludes M32. The dashed black line is the simulated sub-subhalo cumulative distribution of the initial conditions. The solid black line uses the simulated positions of the sub-subhalos at the time when the simulated coordinates of NGC 205 match the observed ones.}
\label{fig:cumd}
\end{figure}

We used a Kolmogorov-Smirnov (KS) test to compare the cumulative distributions of the observed satellites and simulated sub-subhalos in the $X$-direction. These distributions are plotted for the reference simulation in Fig~\ref{fig:cumd} at the time when the simulated position of NGC 205 matches the observed $Y$ and $V_Z$  values (close to pericentre). The red line is the observed cumulative distribution of the 15 satellites (thus including AND XIII and AND XXVII, but excluding M32). Employing the aforementioned exclusion criteria for sub-subhalos close to M31, we show the solid black line which is the simulated cumulative distribution for the reference parameters at the time corresponding to Fig~\ref{fig:xya} - near pericentre. The dashed black line is the distribution at the beginning of the simulation (i.e. corresponding to Fig~\ref{fig:xyb}). The maximum difference between the solid red and black cumulative distributions, which defines the KS-test statistic, is approximately 0.165. The KS-test statistic for the initial distribution of sub-subhalos (dashed black line in Fig~\ref{fig:cumd}) is $\approx$ 0.32, so clearly the tidal encounter efficiently redistributes them.

We can reject the null hypothesis that the original sub-subhalo distribution and the observed satellites come from the same parent distribution at $>99.9\%$ confidence. At pericentre the KS-test statistic is 0.17, but in order to reject the null hypothesis with merely 80\% confidence it would have to be higher than 0.26.

\subsubsection{Monte Carlo sampling technique}
\begin{figure}
\includegraphics[width=8.5cm]{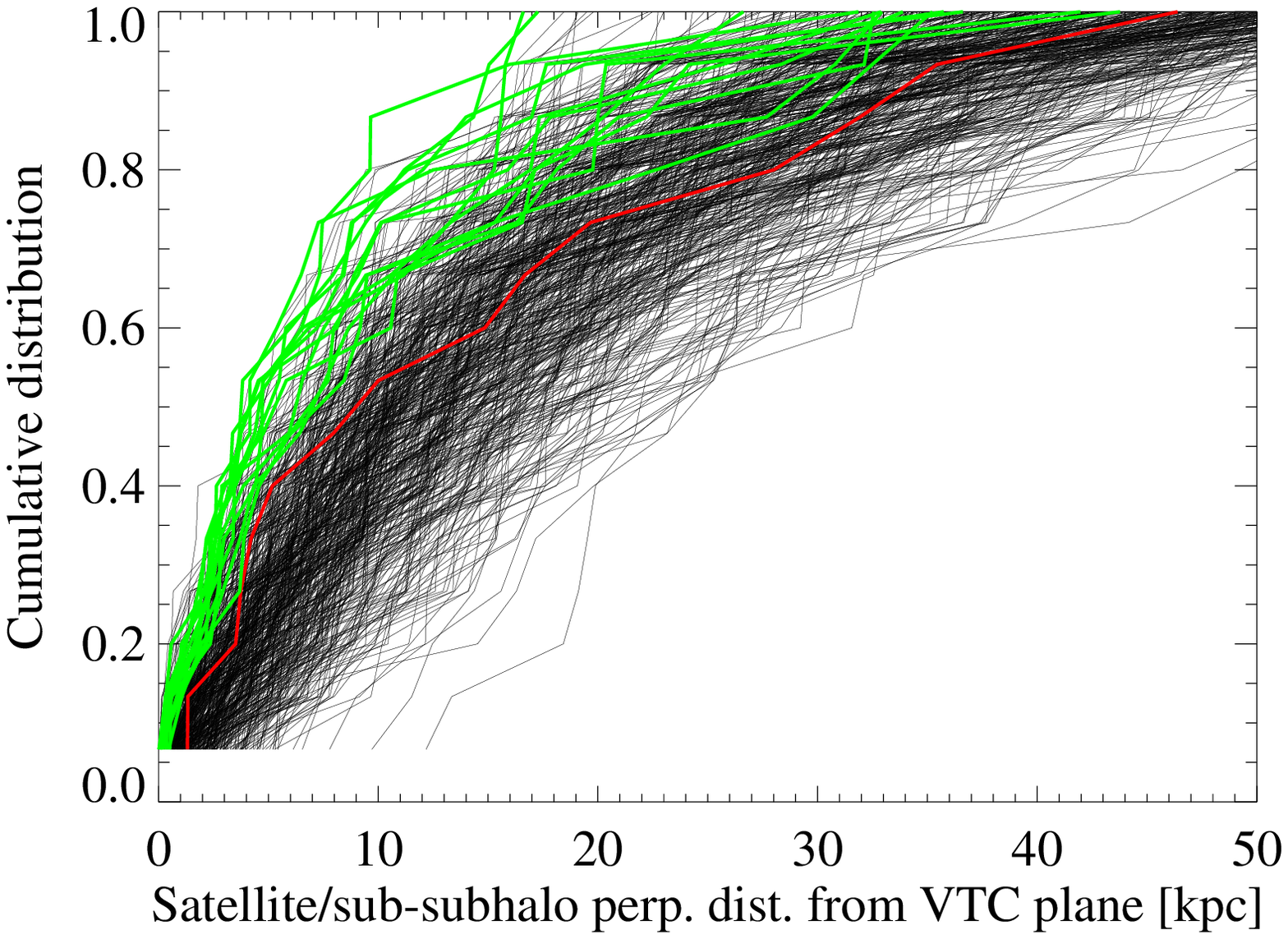}
\caption{Cumulative distribution of distances, $|X|$, perpendicular to the VTC plane for our reference simulation. The solid red line represents the 15 observed satellites (including AND~XIII \& XXVII, but excluding M32). Each of the 1000 black lines represents the cumulative distribution of 15 randomly sampled simulated sub-subhalos from a snapshot near pericentre. The green lines are sampled distributions of 15 sub-subhalos that never cross the red line and are thus thinner than the observed satellite plane.}
\label{fig:boots}
\end{figure}

\protect\label{sec:boot}
Given the distribution of the sub-subhalos near pericentre, the frequency with which we would draw 15 sub-subhalos (corresponding to the 13 co-rotating and 2 counter-rotating) with perpendicular distances from the VTC plane, $|X|$, that are lower than the 15 observed satellites can be estimated. For example, if we only had two observed satellites at 20 and 60~kpc, then a random draw would be successful as long as the $|X|$ of at least one sub-subhalo was below 20~kpc and the other was below 60~kpc. In Fig~\ref{fig:boots} an example of this is given where, over the top of the observed cumulative distribution (red line) of satellite distances, $|X|$, perpendicular to the VTC plane, we plot the cumulative distribution of 1000 randomly drawn sets of 15 sub-subhalos from our reference simulation. A small fraction of these samples (which are given green lines) have thinner distributions than the observed satellites where the green line never crosses the red line. Sampling the simulated sub-subhalo distribution 500k times, we estimated that a distribution thinner than the observed one is found $\sim$1\% of the time for the reference simulation. 

As an aside, here we use NGC 205 as the dominant galaxy in the accreted subhalo, but in principle M32 could be used instead. The probabilities of successfully reproducing the observations with M32 are typically an order of magnitude lower than with NGC~205.

In \S\ref{sec:ks2} we investigate the influence of tweaking various parameters on the derived probability.

\subsubsection{Variation of probability/KS-test statistic along the orbit}
\begin{figure}
\includegraphics[width=8.5cm]{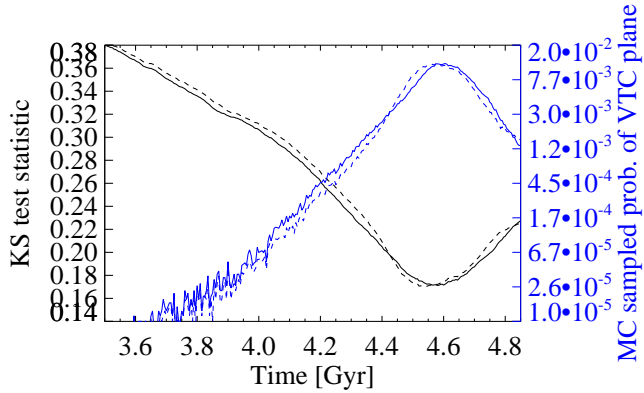}
\caption{The black lines, which relate to the left hand $y$-axis, are the Kolmogorov-Smirnov statistics comparing the simulated and observed distances, $|X|$, perpendicular to the VTC plane (i.e. the maximum difference between the solid black and red lines from Fig~\ref{fig:cumd}) as a function of time for our reference simulation around pericentre. The blue lines correspond to the right hand $y$-axis, which is log-scale, and are the Monte Carlo sampled probabilities of producing a thinner distribution than the observed satellites from the instantaneous distribution of simulated sub-subhalos. The solid lines refer to scenario 1 and the dashed lines to scenario 2.}
\label{fig:ksprob}
\end{figure}

In Fig~\ref{fig:ksprob} we plot both the variation of the KS-test statistic and the Monte Carlo sampled probability of sampling a thin distribution of sub-subhalos, discussed in \S\ref{sec:ks} \& \ref{sec:boot}, along the orbit for our reference simulation. The $y$-axis on the left hand side displays the KS-test statistic and corresponds to the black lines. The logarithmic $y$-axis on the right hand side displays the Monte Carlo sampled probability and corresponds to the blue lines. Both sets of lines display similar behaviour, as expected. When the KS-test statistic drops, the Monte Carlo sampled probability increases. The KS-test statistic drops from 0.58 to 0.17 at pericentre and then quickly increases. The probability increases from  0 to 0.01 at pericentre. The probability stays above 0.005 for around 300~Myr.

\subsection{General trends and coordinates of sub-subhalos at pericentre}
\begin{figure}
\includegraphics[width=8.5cm]{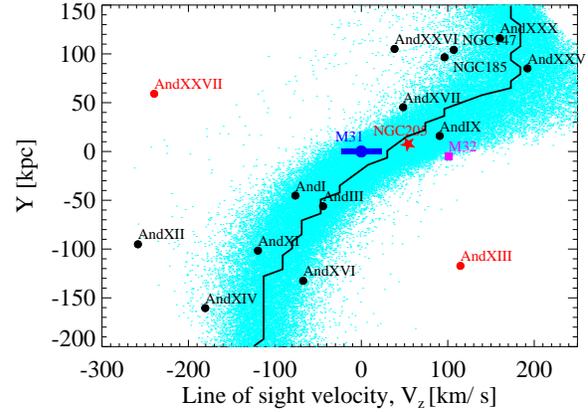}
\caption{The sub-subhalos and observed satellites in the $V_Z$-$Y$ plane (i.e. line of sight velocity relative to M31 and plane of the sky location above or below M31) at the time when the simulated NGC 205's coordinates match the observed ones. The symbols are as per Fig~\ref{fig:xya}. A negative velocity means the object is moving towards the Milky Way. The coordinates of the simulated and observed M31 are identical. The errors in velocity are typically a few $\kms$ and thus are roughly the size of the symbols (McConnachie 2012). The continuous series of points $(V_{z_p}, Y_p)$ discussed in \S\ref{sec:eps} is given as the black line. }
\label{fig:vzy}
\end{figure}

\begin{figure}
\includegraphics[width=8.5cm]{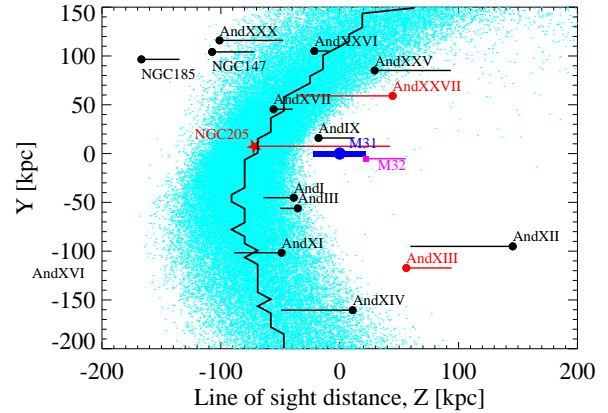}
\caption{The sub-subhalos and observed satellites in the $Z$-$Y$ plane (i.e. line of sight distance and plane of the sky location above or below M31). The symbols are as per Fig~\ref{fig:xya}. The black line emerging from each data point represents the magnitude and direction of the satellite's line of sight velocity vector. The more positive the $Z$ position, the further away the object is from the Milky Way. The coordinates of the simulated and observed M31 are identical. The continuous series of points $(Z_p, Y_p)$ discussed in \S\ref{sec:eps} is given as the black line.}
\label{fig:zy}
\end{figure}

The distribution of distances, $|X|$, perpendicular to the VTC plane is not the only feature that must be reproduced, there is also the distribution of satellites in the $Y$ direction and the $X$-$Y$, $Z$-$Y$ and $V_Z$-$Y$ planes. The distributions of observed satellites and simulated sub-subhalos are shown for the three planes in Figs~\ref{fig:xya}, \ref{fig:vzy} \& \ref{fig:zy}.

In Fig~\ref{fig:vzy} we plot the $Y$-positions of the observed satellites and simulated sub-subhalos against their line of sight velocities, $V_Z$. The vast majority of the observed satellites overlap with the sub-subhalos. The notable exceptions are of course the counter-rotating AND XIII \& XXVII and this is precisely the reason they are not considered to be part of the VTC plane of satellites. Apart from those two, AND XII \& XXVI are in regions of low probability and become increasingly more difficult to account for when using less extended initial distributions of sub-subhalos.

In Fig~\ref{fig:zy} we plot the $Y$-positions of the observed satellites and sub-subhalos against their line of sight distances, $Z$, assuming scenario 2. Recall from Fig~\ref{fig:orbitYZ} that using scenario 1 would lead to NGC 205 being located behind M31 with a positive $Z$. Each observed satellite has a line emerging from the data point, which indicates the direction and magnitude of its line of sight velocity vector. The line of sight distance for each satellite has a great deal more error associated with it than the line of sight velocity, so this is shown more for completeness than for a strict comparison. Clearly several observed satellites have substantial offsets from the turquoise band of simulated sub-subhalos. In fact, AND XVI is not plotted because its coordinates ($Z$, $Y$)=(-307~kpc, -132~kpc) lie off the chart. Nevertheless, these offsets are typically less than 50~kpc, which is around 6\% at that distance ($\sim 780~kpc$ to M31). This is roughly the quoted accuracy of the tip of the red giant branch distance measurement, although using Bayesian techniques \cite{conn1,conn2} claim precisions as low as 2\% for the distances to some of the M31 satellites. The galaxies with larger than 50~kpc offsets are NGC~185, AND XII \& XVI. Interestingly, AND XII is the same galaxy whose line of sight velocity is on the borderline of being too negative, but actually the quoted 1$\sigma$ uncertainty on its distance is 136~kpc.

\subsection{The concentration of sub-subhalos in the $Y$-direction}
\protect\label{sec:ydir}
We saw Fig~\ref{fig:xya} previously and clearly the sub-subhalos are not only narrow in the $X$-direction, but also highly extended in the $Y$-direction - as are the observed satellites. Having said that, the simulated sub-subhalos in Fig~\ref{fig:xya} do appear more concentrated in the $Y$-direction. In particular, we would expect to find more satellites near M31, but perhaps these are difficult to identify against the bright background of M31. Furthermore, the simulated sub-subhalos decrease in density with increasing $|Y|$, but the observed satellites seem to be more prominent at larger $|Y|$. Clearly the significance of these observations depends on the area around M31 that we mask and also the $Y$ limits. In \S\ref{sec:mask} we stated that we excluded all sub-subhalos in an area $\sqrt{X^2+Y^2} < 40 \rm kpc$ around M31, and below $-180$~kpc and above $135$~kpc to exclude the region beyond the PANDAS footprint. However, the precise values of these limits can be argued, although the area around M31 cannot have a radius larger than 50~kpc or it would encroach upon AND I, IX \& XVII. Similarly, $Y> -160$~kpc would exclude AND XIV while $Y<120$~kpc would exclude AND XXX.

\begin{figure*}
\includegraphics[width=17.0cm]{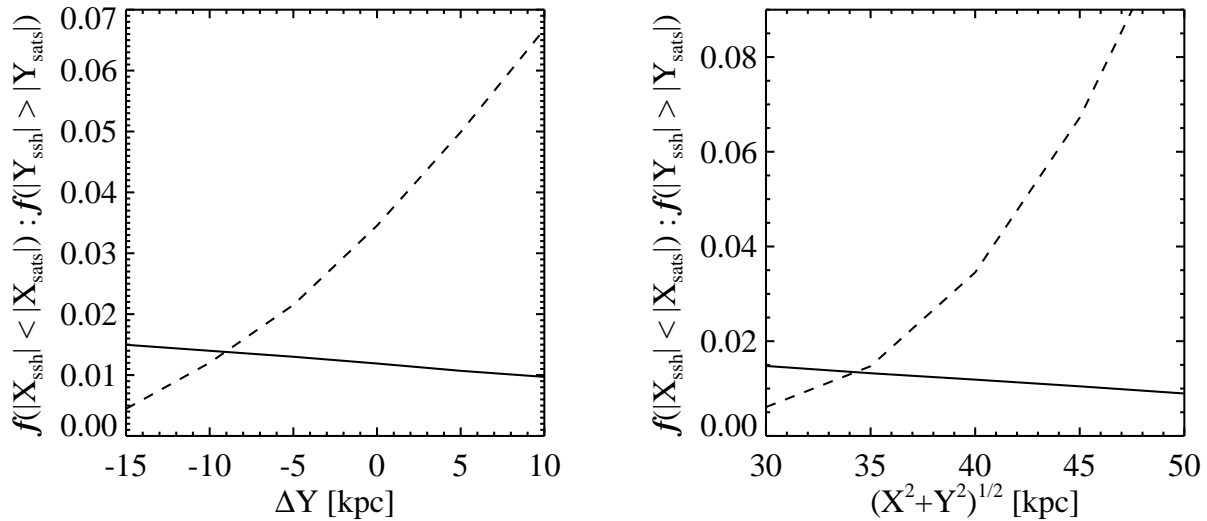}
\caption{The Monte Carlo sampled probability of producing as thin [extended] a distribution of simulated sub-subhalos in the $X$ (solid line)  [$Y$ (dashed line)]-direction as the observed distribution of satellites. We plot these quantities against two variables. In the left hand panel we vary the extent of the sampled distribution of sub-subhalos from the default extent by the value on the $x$ axis. In the right hand panel, we vary the radius of the centrally masked area surrounding M31, where the default radius is 40~kpc.}
\label{fig:varymask}
\end{figure*}

To test the impact of these limits, we computed for the $Y$-direction an analog to the probability of sampling a set of sub-subhalos with a distribution that is thinner than the observed satellites in the $X$-direction. Since it appears difficult to produce as extended a distribution of sub-subhalos in the $Y$-direction, we compute the probability that a sampled set of 15 sub-subhalos is more extended in $Y$ than the observed satellites. Thus, we consider a sample to be successful only if $|Y_{j, \rm ssh}| > |Y_{j, \rm sat}|$ for all $j=1 - 15$. In Fig~\ref{fig:varymask} we plot the successful fraction of 500k random samplings for the $Y$-direction (dashed line) against two variables. The left hand panel varies the upper and lower limits from -15~kpc to +10~kpc, meaning that for -15~kpc we shorten the upper limit from 135~kpc to 120~kpc and the lower limit we shorten from -180~kpc to -165~kpc. We hold the radius of the masked region around M31 at the default 40~kpc. Unsurprisingly, the probability increases the broader we make the sampled area since the number of large $Y$ sub-subhalos increases. The black solid line shows how the probability of producing a thin distribution in the $X$-direction varies when the sampled area is extended. The influence is considerably weaker than in the $Y$ direction.

In the right hand panel we fix the upper and lower $Y$ limits to the default values and vary the radius of the masked region around M31. Again this has a strong influence on the probability of sampling an extended set of sub-subhalos in $Y$, but a fairly weak influence on the chances of generating a thin distribution in $X$. To get a more constrained probability for the $Y$-direction we would need to know conclusively that no satellites are present in the excluded regions. Until that time, the $Y$-distribution is a promising and likely fairly stringent, but unreliable, gauge of the probability of our scenario replicating the observed distribution of satellites.

\subsection{The concentration of sub-subhalos in three dimensions}
\begin{figure}
\includegraphics[width=8.5cm]{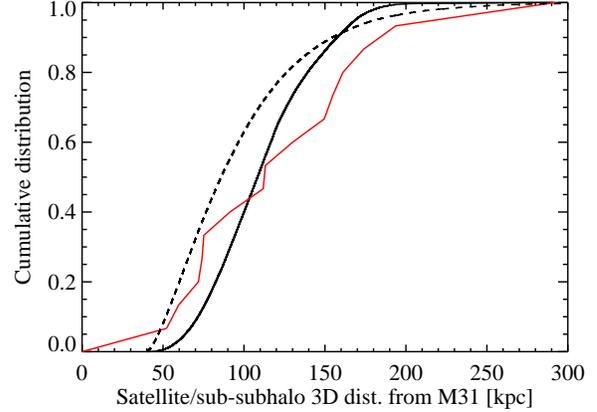}
\caption{Cumulative distribution of the three dimensional distances of the observed satellites from M31 (red line). For comparison we also show the cumulative distribution of the 3D distances for the simulated sub-subhalos near pericentre (solid black line) and for the initial distribution of sub-subhalos (dashed black line). The solid red line includes all 15 observed satellites, including the counter rotating AND~XIII \& XXVII, but excludes M32.}
\label{fig:cum3d}
\end{figure}

\protect\label{sec:rdir}
In Fig~\ref{fig:cum3d} we plot the cumulative distribution of the three dimensional distances of the observed satellites (red line) from M31. We also plot the cumulative distribution of the 3D distances of the sub-subhalos near pericentre from M31 (solid black line) and for the initial distribution (dashed black line). We use the standard exclusion criteria for the projected area around M31.

The KS statistic between the red and black solid lines is 0.2 meaning there is no evidence to suggest the two distributions are different.

\subsection{Trends with parameter variations using a series of simulations}
\protect\label{sec:ks2}

\begin{figure*}
\includegraphics[width=17.0cm]{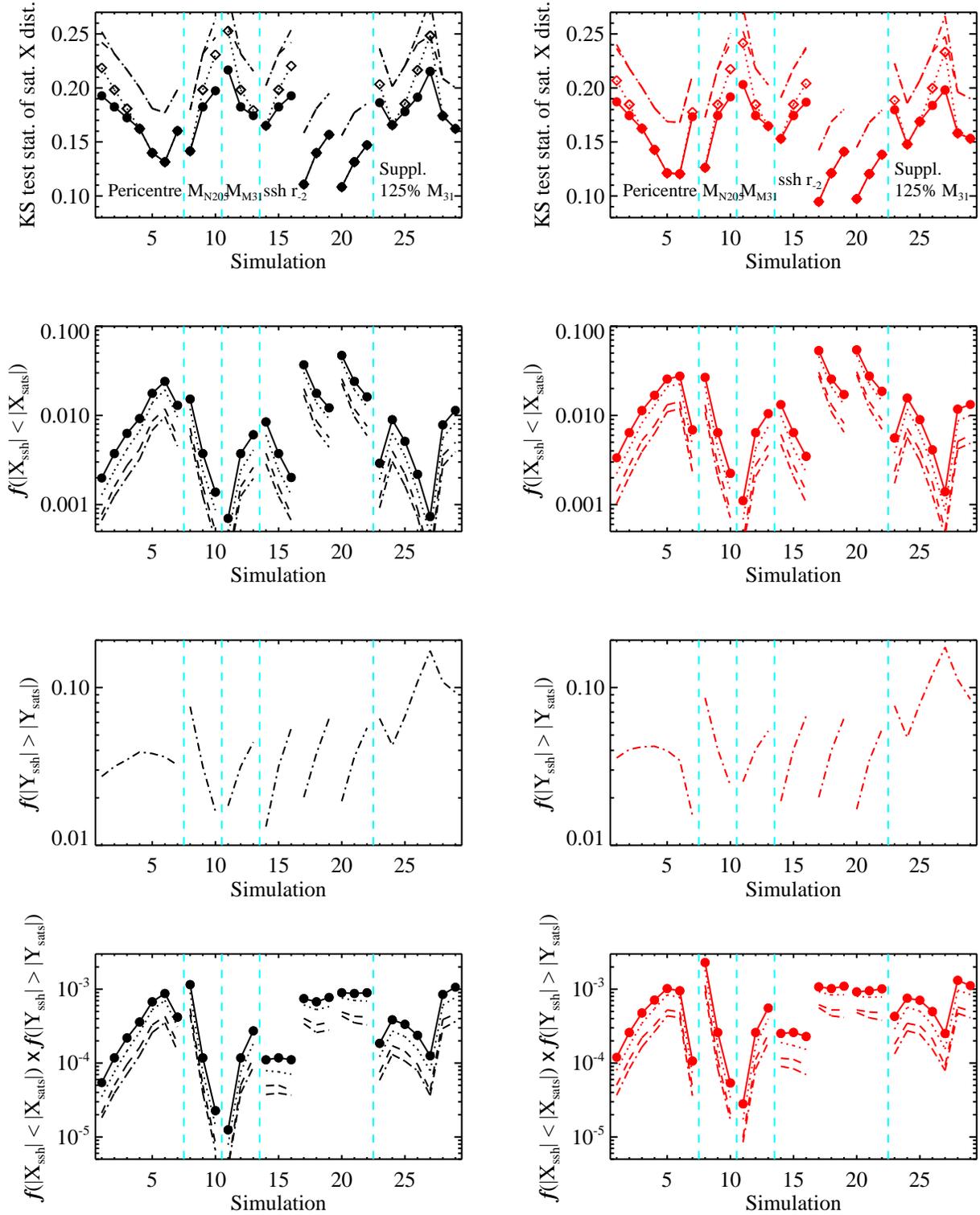}
\caption{These figures show the variations in some parameters across our series of simulations listed in table~\ref{tab:sims}. The left hand panel refers to scenario 1 and the right hand panel to scenario 2 and this is the format for all four rows of figures. The top row shows the results of the KS test which was initially discussed in \S\ref{sec:ks} and then discussed further in \S\ref{sec:ks2}.  The dot-dashed lines refer to all 15 satellites, the dashed lines exclude AND XIII, the dotted lines exclude AND XXVII and the solid lines (with filled circles) use neither of these two counter rotating satellite. Using the same line types, the second row shows the Monte Carlo sampled probability of producing a thinner distribution of sub-subhalos in the $X$-direction than the observed satellite distribution as discussed in \S\ref{sec:boot}. The third row shows the Monte Carlo sampled probability of producing as extended a distribution of sub-subhalos in the $Y$-direction as the observed distribution of satellites, as discussed in \S\ref{sec:ydir}. For this we only plot the dot-dashed line representing all 15 satellites. In the last row we show the product of the Monte Carlo sampled probabilities from rows two and three. The turquoise lines isolate related simulations to show the influence of (typically) a single parameter. 
}
\label{fig:scen2}
\end{figure*}

In addition to these plots of the KS test statistic and the Monte Carlo sampled probability of the thinness of the $X$-distribution (Fig~\ref{fig:ksprob}) with time near pericentre for the reference simulation, we also plot their values (and the probabilities for the extent of the $Y$-distribution) near pericentre for all our simulations in Fig~\ref{fig:scen2}. The details of the different simulations are given in table~\ref{tab:sims}. 

For ease of comparison, we separate the various simulations into blocks delineated by the vertical dashed turquoise lines. The first block is for simulations with the reference parameters (table~\ref{tab:sbpar}) but different initial tangential velocities (i.e. pericentres). We tried seven different values: 19, 24.5, 31, 37, 52, 72 and 126~kpc (25, 30, 35, 40, 50, 60 and 80~$\kms$; simulations 1-7). For these simulations, the values for the mass of M31, NGC 205 and the scaling of the sub-subhalos distribution were exactly as per the reference parameters in table~\ref{tab:sbpar}. The second and third blocks have constant initial tangential velocity of $30~\kms$ and use the reference parameters, except for varying (2) the mass of NGC205 (30, 100 and 150\% of the reference value; simulations 8-10 respectively) and (3) the mass of M31 (80, 100 \& 150\% of the reference value; simulations 11-13 respectively). The fourth block varies only the radial scalings for the sub-subhalo distribution, $r_{-2}$, from 80, 100 and 120\% of the reference value for three different initial velocities (30, 50 and 60~$\kms$ - simulations 14-16, 17-19 and 20-22 respectively). The last block is for simulations where the mass of M31 is always scaled to 125\% of the reference value. Amongst them, simulations 24-27 use a fixed initial tangential velocity of $35~\kms$ (pericentre of 27.5~kpc) and then scalings for the sub-subhalos distribution of 100\%, 120\%, 150\% and 200\%.  Scenario 23 has an initial tangential velocity of 30~$\kms$ (corresponding to a pericentre of 22.5~kpc) and a sub-subhalo distribution inflated to 120\% of the reference value. Scenarios 28-29 use initial tangential velocities of 50 and 65~$\kms$ (corresponding to pericentres of 46.5 and 72~kpc respectively) and sub-subhalo distributions inflated to 150\% of the original. 

For the KS test statistic of the $X$-distribution in the first row of Fig~\ref{fig:scen2}, and all subsequent rows of figures, we make separate plots for scenarios 1 and 2 (left and right hand panels respectively). Scenario 1 is always plotted in black and scenario 2 in red. The different line types distinguish between the different number of observed satellites that we considered. The dot-dashed line considers all 15 satellites, the dashed line excludes AND XIII, the dotted line excludes AND XXVII and the solid line excludes both AND XIII \& AND XXVII (which are the two counter rotating satellites). Scenario 2 gives slightly lower KS test statistics than scenario 1, but the trends with variables are the same: there is a preference for lower NGC 205 masses, higher M31 masses and a tighter distribution of sub-subhalos. 
The Monte Carlo sampling of the $X$ distribution of satellites, which is plotted in the second row of Fig~\ref{fig:scen2}, exactly reflects the KS test - lower KS test statistics lead to larger probabilities. The ideal initial tangential velocity, when using the reference mass for M31, appears to be around $60~\kms$ which gives a probability of 0.01. Regardless of the parameters we investigated here, the probability of creating such a thin distribution of satellites (ignoring the other dimensions) is at best a one in thirty chance.

The third row of Fig~\ref{fig:scen2} gives the probabilities of producing as extended a distribution of sub-subhalos in the $Y$-direction as the observed satellites. Here we use the default exclusion limits as discussed in \S\ref{sec:mask} and \S\ref{sec:ydir}. We only plot a single line-type corresponding to all 15 satellites since there is not a significant impact on the probability from the counter-rotating satellites. In the fourth row of Fig~\ref{fig:scen2} we give the product of the probability of producing a thin distribution in the $X$-direction and an extended distribution in the $Y$-direction (rows 2 and 3 respectively). Velocities around 50-60~$\kms$, low masses for NGC 205 and high masses for M31 are strongly preferred. Interestingly, the concentration of the sub-subhalo distribution does not appear to play a role.

\subsection{Monte Carlo sampling in more than one dimension}
\protect\label{sec:mc2}

\begin{figure*}
\includegraphics[width=17.0cm]{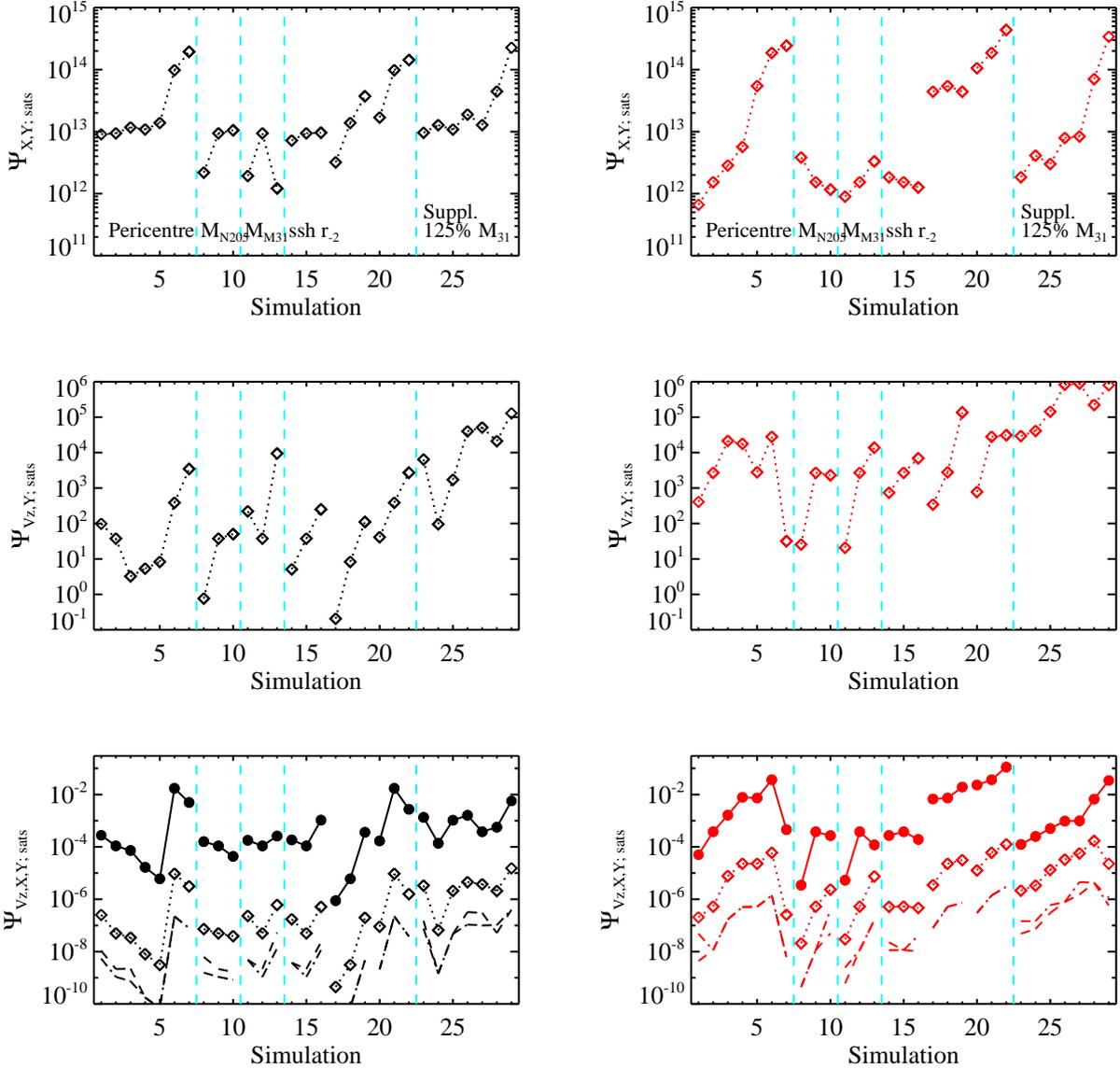}
\caption{As per Fig~\ref{fig:scen2} these figures show the variations in some parameters across our series of simulations listed in table~\ref{tab:sims}. The left hand panels refer to scenario 1 and the right hand to scenario 2.  The first and second rows show the variation in $\Psi_{X,Y;\rm sats}$ and $\Psi_{V_Z,Y;\rm sats}$ respectively for the 13 co-rotating satellites as discussed in \S\ref{sec:mc2}. The third row shows $\Psi_{V_Z,X,Y;\rm sats}$ for the 11 core satellites (solid line with filled circles), the 13 co-rotating satellites (dotted line with empty diamonds), 14 satellites excluding AND XXVII (dashed) \& 14 satellites excluding AND XIII (dot-dashed). The turquoise lines isolate related simulations to show the influence of (typically) a single parameter.}
\label{fig:scen3}
\end{figure*}

To compliment these illustrative probabilities we made more concrete Monte Carlo samplings of the possibility that the distribution of observed satellites comes from the sub-subhalo distribution near pericentre. We do this for the $X$-$Y$ and $V_Z$-$Y$ planes by initially calculating the number of sub-subhalos within a small circle around each satellite in the plane divided by the total number of sub-subhalos in the observed area (see \S\ref{sec:mask} for a discussion of the excluded regions). For the $X$-$Y$ distribution this is simply the number within a radius of 10~kpc around the $j^{\rm th}$ observed satellite and we refer to this as $n_j^{\rm X, Y}$. For the $V_Z$-$Y$ plane we also use a ``radius" of $\lambda_{V_Z,Y}=10 \rm kpc \kms$. We also do this for the 3D distribution $X$, $Y$ and $V_Z$. We then take the product of each count separately over the selected number of satellites, such that $\Psi_{X,Y; \rm sats}=\Pi_{j=1}^{N_{\rm sats}}n_j^{\rm X, Y}$, $\Psi_{V_Z,Y; \rm sats}=\Pi_{j=1}^{N_{\rm sats}}n_j^{\rm V_Z, Y}$ and $\Psi_{V_Z,X,Y; \rm sats}=\Pi_{j=1}^{N_{\rm sats}}n_j^{\rm V_Z,X, Y}$. In the three rows comprising Fig~\ref{fig:scen3} we plot respectively the values for $\Psi_{X,Y; \rm sats}$, $\Psi_{V_Z,Y; \rm sats}$ and $\Psi_{V_Z,X,Y; \rm sats}$ for all simulations, with a constant, but arbitrary numerical scaling. For $\Psi_{X,Y; \rm sats}$ and$\Psi_{V_Z,Y; \rm sats}$ we only plot the values for the 13 co-rotating satellites to avoid clutter. It is worth noting that the typical value one finds for sampling the 13 satellites from a typical {\it subhalo} distribution around an M31 sized galaxy is around 1000 for  $\Psi_{V_Z,Y; \rm sats}$ and $10^{-12}$ for  $\Psi_{V_Z,X,Y; \rm sats}$, whereas values $>10^4$ and $>10^{-5}$ are easily achievable for the {\it sub-subhalos} at pericentre. This simply means the sub-subhalo scenario is intrinsically more likely. 

For the bottom two rows of Fig~\ref{fig:scen3}  there are large differences between scenario 1 and 2 for the preferred initial tangential velocity. The key tension with the preferred parameters for only producing a thin disk of satellites, is the scaling of the sub-subhalo distribution. This arises because the observed satellites are relatively spread out around the trajectory of the turquoise dots in Fig~\ref{fig:vzy} and therefore a few satellites can only be surrounded by a sizeable density of sub-subhalos if the sub-subhalos are spread out more.

\subsubsection{Probabilities from Monte Carlo sampling in more than one dimension}
\begin{figure*}
\includegraphics[width=17.0cm]{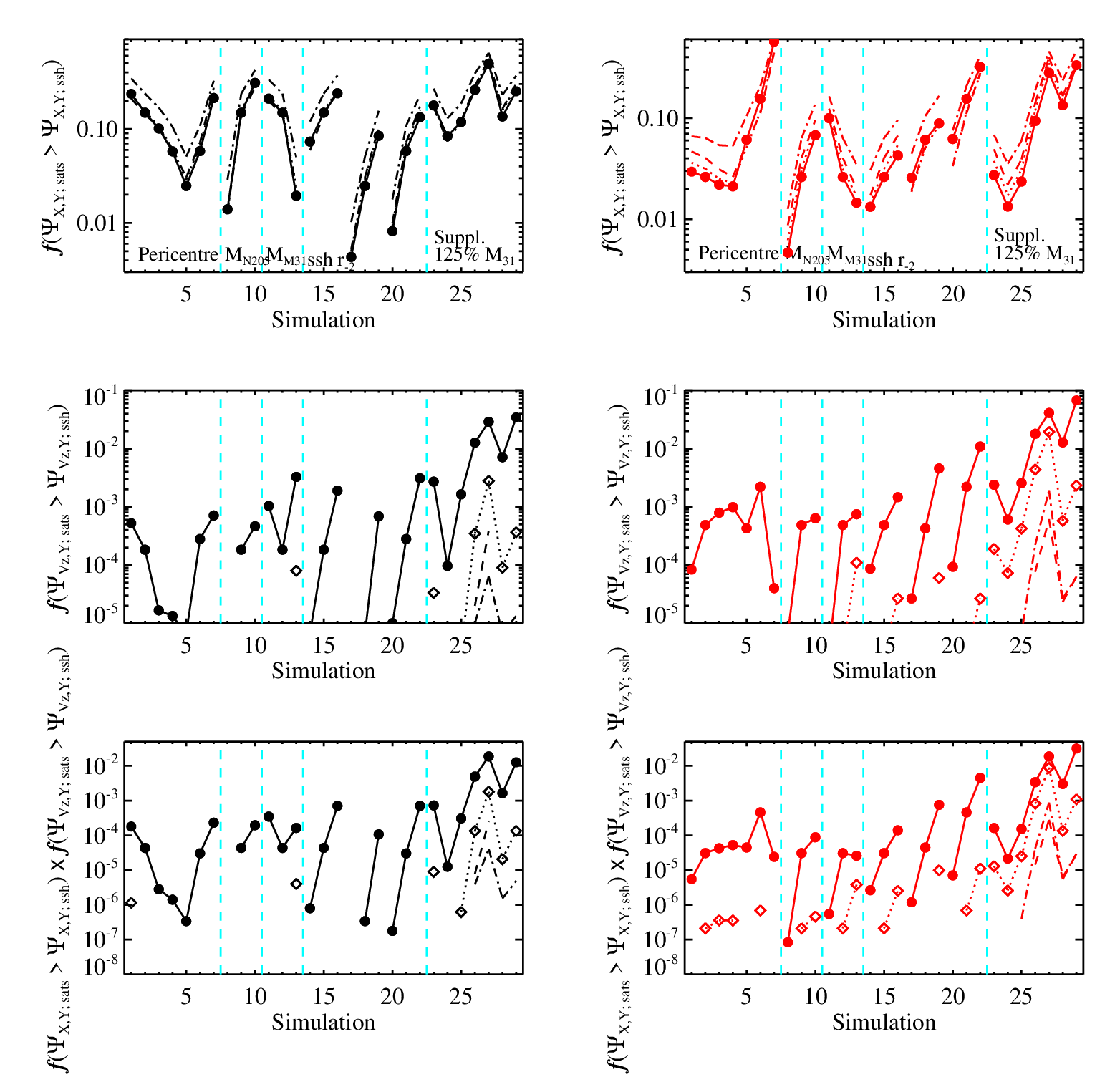}
\caption{As per Fig~\ref{fig:scen2} these figures show the variations in some parameters across our series of simulations listed in table~\ref{tab:sims}. The left hand panels refer to scenario 1 and the right hand to scenario 2. The first and second rows show the Monte Carlo sampled probability of producing the 2D distribution of satellites in the $X$-$Y$ plane and $V_Z$-$Y$ plane respectively, as discussed in \S\ref{sec:pmc2}. The solid lines (with filled circles) represent the 11 core satellites, the dotted lines (with empty diamonds) represent the 13 co-rotating satellites, the dashed lines refer to 14 satellites excluding AND XXVII \& the dot-dashed lines refer to 14 satellites excluding AND XIII. The final row shows the product of the probabilities from the top two rows. The turquoise lines isolate related simulations to show the influence of (typically) a single parameter.}
\label{fig:scen4}
\end{figure*}

\protect\label{sec:pmc2}
We then repeated this analysis, but instead of calculating the number of sub-subhalos around the selected number of observed satellites, we randomly selected the required number of satellites from the sub-subhalo distribution to represent the satellites.  These statistical distributions we refer to as $\Psi_{X,Y; \rm ssh}$ and $\Psi_{V_Z, Y; \rm ssh}$, where ssh stands for sub-subhalos.  We do not use $\Psi_{V_Z,X,Y; \rm ssh}$ because there are insufficient sub-subhalos to sample from, but for the other two we compute $3\times10^5$ realisations with different random sets of sub-subhalos representing the observed satellites. We quote the probability as the fraction of samplings for which $\Psi_{\rm sats}>\Psi_{\rm ssh}$ and we plot this for the $X$-$Y$ plane and the $V_Z$-$Y$ plane in Fig~\ref{fig:scen4}. The dot-dashed lines include 14 satellites except for AND XIII, the dashed line only excludes AND XXVII and the dotted line excludes both counter rotating satellites (AND XIII \& AND XXVII). Although there is no reasonable justification for doing so, the solid line excludes both counter rotating satellites and the two main outliers from the $V_Z$-$Y$ satellite distribution (AND XII \& AND XXVI).

The $X$-$Y$ plane only weakly constrains the likelihood of the heights above the plane, but rather the 2D distribution of satellites in the plane of the sky. Here the probabilities are not adversely affected by the counter-rotating satellites. The probability increases with higher NGC 205 masses, lower M31 masses and larger radial scalings of the sub-subhalos distribution which is the exact opposite of what was found for the $X$ distribution. The trend of probability with initial tangential velocity changes from scenario 1 to 2, but velocities $50~\kms$ seem preferred for scenario 2.

For the $V_Z$-$Y$ plane the key difference is that a larger mass for M31 is favoured. There is not too much variation in probability with initial tangential velocity for scenario 2, although $60~\kms$ seems ideal. For scenario 1 velocities less than $60~\kms$ and greater than $30~\kms$ are disfavoured. It appears the most important parameter for this plane is simply the radial scaling of the sub-subhalo distribution: see simulations 24-27 where the scaling goes from 100\% -200\% of the original size; and 14-16, 17-19 \& 20-22 where it goes from 80\% - 120\% for three different initial tangential velocities. The main reason for this is that the spreading out of the sub-subhalos allows the distribution to overlap with the dispersion of satellites in this plane which produces some moderate outliers (AND XII \& AND XXVI). The probabilities for scenario 2 are generally slightly larger than scenario 1. A few simulations in scenario 1 have no measurable probability of even generating the 11 core satellites and very few simulations are able to match the observations well enough to have measurable probabilities when trying to account for 13 satellites. It seems this is only possible for larger masses of M31 and more extended sub-subhalo distributions. In fact, it seems like increasing the mass of M31 to 125\% of the reference mass for simulations 23-29 (and 150\% for simulation 13) is the key parameter for producing a better match to all 13 co-rotating observed satellites. This follows because a larger mass for M31 leads to a larger tidally induced spread in sub-subhalo velocities.

An interesting result is that decreasing the radial extent of the sub-subhalos and the mass of NGC 205 leads to increases in the probability of generating a very thin distribution of satellites, however it also leads to dramatic decreases in the probability of reproducing the satellites in the $V_Z$-$Y$ plane. 

In the last row we combine the two probabilities from the first two rows. This gives no consideration of the $Z$-$Y$ positions, but the combination of measurement uncertainties for the line of sight distances and the general spread in the simulated sub-subhalo distribution should make this a relatively minor correction. These two panels in the last row show a clear preference for certain initial tangential velocities, with $60~\kms$ standing out for scenario 2. It also hints at a threshold mass for both M31 and NGC 205 below which the probability is negligible. Probably the key factor is still the scaling of the sub-subhalo distribution with extended distributions preferred.

\subsection{Unmodelled factors}

The result of these probability studies is that despite the simulations broadly agreeing with the general patterns in the phase space distribution of the satellites, the probabilities recovered suggest the situation is more intricate than our simple model. It is important to bear in mind that the scenario is generally plausible, meaning that it is possible to sample as thin a distribution of sub-subhalos and the observed distribution of line of sight velocities and positions, but the resulting probabilities are low. We discuss below possible factors that may complicate the scenario.

The radial extent of the distribution of sub-subhalos is not necessarily a simple scaled-down version of the subhalo distribution around M31 (or the Milky Way). Tidal forces from M31 might truncate the sub-subhalo distribution at high redshifts and this more concentrated distribution would change the probabilities. Alternatively, our choice of $r_{200}$ and also $r_{-2}=0.81r_{200}$ might underestimate the extent of the sub-subhalo distribution insofar as the probability of star formation in a sub-subhalo might be linked to that sub-subhalo's orbit. Probably a more significant factor is just that the radial distribution of sub-subhalos is stochastic and varies significantly from subhalo to subhalo and on top of that is anisotropic to some degree. 

There is also the possibility that in-transit encounters have affected the distribution of sub-subhalos. In particular, a scattering from M33 might have changed the orbit and also disturbed the sub-subhalo distribution meaning that a smooth distribution like we present here would struggle account for the resulting orbits. In the same vein, as mentioned in \S\ref{sec:lumfunc} the two larger satellites NGC 147 and NGC 185 are too massive to be members of the subhalo group of NGC 205. The most obvious reason for their proximity is that they became loosely bound to the group. In principle their interaction with the group could also cause some minor variations in the velocities of the other satellites making them less ordered than our simulations suggest. In addition, M32's presence might have caused similar deviations from our predictions.

Another factor that has gained much attention recently is that any peculiarities in the Local Group, and deviations from expectations derived from simulations, might arise because the Local Group is composed of a pair of massive galaxies in close proximity. Many recent simulations like \citealt{elvis,mcalpine15} make sure to focus only on areas of cosmological simulations with Local Group analogs when making comparisons between observations of the Milky Way or M31 and simulated halos. Several papers (for example \citealt{knebe11,teyssier12,fouquet12}) look specifically at the exchange of satellites between M31 and the MW. There is clearly a continual interaction between these two galaxies. Nevertheless, this probably has little influence on the detailed probabilities of our scenario, but it could have important implications for the frequency of the interloping, counter-rotating satellites.

Most of these factors, in particular the initial, anisotropic distribution of sub-subhalos and the influence of in-transit encounters, would be automatically taken into account in cosmological simulations and that is clearly where more concrete answers must be sought.

\subsubsection{Proper motions}
The line of sight velocities of the simulated sub-subhalos are relatively small for objects near pericentre that have orbited from a large distance. This is of course because the motion is mostly in the plane of the sky. For the initial velocities (pericentres) of 25, 30, 35, 40, 50, 60 and 80~$\kms$ (19, 24.5, 31, 37, 52, 72 and 126~kpc) the motions in the plane of the sky at pericentre are roughly 530, 500, 460, 435, 390, 350 and 270~$\kms$ respectively. If measured, these large proper motions could provide a strong constraint on the scenario. Away from pericentre, these velocities would drop quickly meaning that the proper motions of the satellites, relative to M31, should decrease with $Y$ coordinate and NGC 205 should have the largest value. Such large orbital velocities are not unheard of in the Local Group. The Magellanic Clouds, in particular the Large Magellanic Cloud at $>300~\kms$, have very large tangential velocities relative to the Milky Way (\citealt{kallivayalil06a,kallivayalil06b,piatek08,kallivayalil13}).

\subsubsection{Galaxy associations}
\protect\label{sec:galass}
It has been argued that observed galaxy associations are too sparse or loose to explain the VTC plane of satellites. The absolute V-band magnitudes of NGC 205 and the brightest satellite in the co-rotating thin plane, AND I, are -16.5 and -11.9 - a difference of 4.6. \cite{tully06} presented the members of several galaxy associations that are independent of both each other and of Andromeda. NGC 3109 has a companion galaxy with B-band magnitude difference of 6.2 and projected distance of 25~kpc. According to \cite{sand15}, it has a second dwarf companion at $\sim$72~kpc. Of the remaining four groups with satellites that are more than 3 magnitudes dimmer than their host, three have companions with projected distances less than 45 kpc (32, 40.5 and 43.5 kpc) and magnitude differences between 5.1 and 7. One system does not have an observed nearby companion. Since these companion galaxies are typically at the observable threshold, it seems perfectly plausible that other, dimmer, nearby companions are present.

A common question is the location of these groups around M31 (or the Milky Way), since others must have fallen in. The first point is there are only a handful of LMC/SMC or NGC 205/M32 sized halos expected from simulations. From the Aquarius project there are 5 halos more massive than $10^{10}\msun$ and another 7 more massive than $10^{9}\msun$. Of those, the earlier accreted subhalos might not have had a distribution of bright sub-subhalos due to their proximity to M31. Furthermore, there is clearly not a bounty of SMC/NGC 205 sized halos being accreted given that after the SMC and LMC, the Milky Way has no massive nearby satellites. 

Another related question, which we plan to investigate, is the survival of other disks of satellites. In particular, how long does it take to tidally destroy all satellites after infall?

\section{Discussion and Conclusions}
\protect\label{sec:conc}
In this article we explored the idea that Andromeda's vast, thin plane of co-rotating satellites is formed by the tidal break-up of a subhalo group during a close pericentric passage of M31. NGC 205 was assumed to be the galaxy at the centre of the subhalo and it was initially surrounded by many bright sub-subhalos. This is consistent with the observations of \cite{tully06} who found the majority of independent nearby groups have a host and bright satellite (comparable to NGC~205 and AND I) within a projected distance of 25 and 45~kpc of each other. Deeper imaging is required to see if the dimmer satellites are present (\citealt{sand15}).

We showed that the observed luminosity function of the subhalo group, which we propose dispersed into the vast thin plane of satellites, is compatible with the mass function of sub-subhalos from the simulations of \cite{springel08}. Therefore, an advantage of this subhalo group scenario is that it could go some way to mitigating the TBTF problem, since only sub-subhalos/satellites that are part of certain subhalo groups are lit up.

Using standard parameters, we made N-body models for M31, NGC 205 and 200k point particles representing the sub-subhalos which initially orbit NGC 205. NGC 205 and the sub-subhalos initially began from a distance of around two times M31's virial radius and orbited towards M31 with a range of different initial velocities tangential to M31. The initial tangential velocity range was fairly typical of subhalo velocities from cosmological simulations. During the pericentric passage, the sub-subhalos were spread out in the orbital direction, but they were compressed in the transverse direction. This greatly improved the visual agreement with the observed distribution of satellites. 

To compare the distribution of sub-subhalos with that of the observed satellites in the $X$-direction (the distance perpendicular to the vast thin plane) we used a KS test and a Monte Carlo sampling method. For these tests we used all 15 observed satellites including the 13 members of the VTC plane of satellites and the 2 counter-rotating satellites. We made all comparisons when the simulated location and line of sight velocity of NGC 205 agreed with the observed ones. Specifically, $X$, $Y$ and $V_Z$ were identical, but the line of sight distance $Z$ from M31 depended on the initial tangential velocity. 

We performed a series of simulations with different initial conditions facilitated by changing the initial tangential velocity (to give a different pericentre), the masses of NGC 205 or M31 and also the radial extent of the sub-subhalo distribution. The lowest KS-test statistics were found near pericentre, which NGC 205 is currently close to, and for the majority of our simulation parameter-space the null hypothesis, that the sub-subhalo distribution and the observed satellites come from the same parent distribution, could not be rejected at any reasonable confidence level. The Monte Carlo sampling method to determine the probability of drawing a distribution of sub-subhalos from our simulation that is thinner than the observed satellite plane gave a probability that varied significantly with the radial extent of the sub-subhalos, the pericentre and the mass of NGC~205's dark matter halo. Using the reference distribution of sub-subhalos and mass for NGC 205 we found the highest probability of producing a plane of satellites as thin as the observed 15 satellites was around 1\%.
 
Although the thin distribution of observed satellites in the $X$-direction is the most apparent concern, the highly extended distribution of observed satellites in the plane of the sky ($Y$-direction) is also troubling. To this end, we made a similar calculation of the probability of producing as extended a distribution satellites in the plane of the sky, but found that this depended critically on the size of the area that is excluded from sampling due to M31 saturating its surroundings and the chosen outer limits of the sampled region.

We also calculated the probability of agreement between the observed satellites and our simulations with Monte Carlo sampling of two combinations (1) the observed line of sight velocities with the $Y$ distribution, and (2) the $X$-$Y$ plane (the plane of the sky). The $X$-$Y$ easily gave probabilities greater than 0.1 for many sets of simulation parameters. The $V_Z$-$Y$ plane showed very good visual agreement, bolstered by a self-devised quantity for the likelihood of the satellites relative to the simulated sub-subhalos, discussed in \S\ref{sec:eps}. However, the details of the satellite phase space distribution made a convincing match difficult. Ignoring the two counter-rotating satellites, the probability is still very low - at best around $10^{-4}$ unless the radial extent of the sub-subhalos is inflated by 50\% or more. Such a tactic can increase the probability by a factor of 10 or more, but this in turn reduces the probability of randomly finding 15 satellites with very small distances perpendicular to the VTC plane. Ignoring two other marginal outliers (leaving 11 satellites) improves the best probabilities to $10^{-3}$ for the reference scaling of the sub-subhalo distribution and up to $10^{-2}$ for a mere 20\% increase in that parameter.  If those two outliers must be included then a slightly larger mass (at least 125\% of the reference value) for M31 is required. Although the probabilities we find for this scenario are seemingly low, they are still competitive with other models.

We argue that this may simply have to do with our model not being sophisticated enough to include potential encounters with other galaxies along the orbit, such as M33, or simply M32, NGC 147 and NGC 185. Such encounters would inject a random component into the phase space distribution of sub-subhalos. Other complications might be that certain satellites are simply interlopers, or the initial sub-subhalo distribution is slightly anisotropic. Another factor that would increase the probability would be if there are as yet undiscovered satellites close to M31, within our excluded region of 40~kpc. We expect around 10 such undetected dwarf galaxies.

Thus this scenario seems like an interesting candidate for explaining the bulk of the vast thin plane of co-rotating satellites and cannot yet be ruled out. There are still complexities that might be very important that are not accounted for here and really require high spatial and temporal resolution cosmological simulations such as the Caterpillar suite (\citealt{griffen16}) and ELVIS suite (Exploring the Local Volume in Simulations; \citealt{elvis,mcalpine15}), the latter of which crucially focuses on binary Local Group analogs.

It will also be interesting to see how the scenario plays out for the Milky Way satellite system, the detected plane of satellites around Centaurus A (\citealt{tully15}), as well as a second plane of satellites in Andromeda (\citealt{shaya13}). To explain the second plane in Andromeda there would need to be a separate encounter between a subhalo group of galaxies and M31. This implies that all low mass satellite galaxies form in sub-subhalos and their locations are caused by encounters with the host that separated them from their subhalo host.

To close we re-emphasise the various assumptions on which this scenario rests, which are also predictions.

\begin{enumerate}[(i)]
\item Halos like the Milky Way, M31, LMC or M33 with $M_{\rm vir} > 10^{11}\msun$ inhibit the star formation in their own nearby subhalos, or tidally disrupt them. 
\item Only subhalos that fell from (or at least orbited to) well outside M31's virial radius formed large quantities of stars. 
\item There is a subhalo mass window, $M_{\rm vir} \sim 10^{9-10}\msun$, for which basically all bound sub-subhalos formed significant numbers of stars.
\item The vast majority of galaxies with $M_{\rm vir} < 10^{8}\msun$ do not form in isolation, but within the sub-subhalos of a subhalo group.
\item Every separate plane of satellites is caused by a distinct encounter between a subhalo group and a host galaxy.
\item M32, AND XIII \& XXVII are interlopers of the VTC plane of satellites.
\item There should be around 10 undetected satellites within 40~kpc of M31.
\item NGC 147 \& 185, which are gravitationally bound to each other, were also dragged by (or led) NGC 205 towards M31.
\item Field dwarf ellipticals with $M_{\rm vir} \sim 10^{9-10}\msun$ should have observable dwarf satellites with the same luminosity function and spatial distribution as discussed here (following \citealt{springel08}).
\item For reasonable parameters, and ignoring the other coordinates, the chance of producing as thin a distribution of satellites as the observed VTC plane at pericentre is less than roughly 1\%, therefore such ultra-thin distributions should not be found around a statistically significant proportion of Andromeda sized galaxies.
\item The proper motions of NGC 205 and the dwarf galaxies comprising the VTC plane of satellites should be large. For example, NGC 205 should have a plane of the sky motion of at least 300~$\kms$ relative to M31. This was also predicted by \cite{howley08} to explain NGC 205's tidal features.

\end{enumerate}

\begin{table*}
\begin{center}
\begin{tabular}{|c|cccc|cc|cc|cccc|}
\hline
Parameter &$M_{\rm DM}$ &$\rho_s$				&$r_s$& $r_{\rm 200}$&$M_{\star}$&$h_{R,\star}$&$M_{g}$&$h_{R,g}$&$M_{\rm b}$ &$\Sigma_e$&$R_e$&$n_s$\\
Units& $10^{10}\msun$&$10^6\msun kpc^{-3}$&kpc  & kpc&$10^8\msun$& kpc	 &$10^8\msun$&kpc&$10^8\msun$&$10^4\msun kpc^{-2}$			 &kpc & \\
\hline
M31         & 178 &15 & 15.4 & 185.0 & 600 & 5.9 & 80 & 5.9 & 377&12.5 & 1.93 & 1.71 \\
NGC 205 & 1.8& 1.1 & 8.38   &  60.0   & 4  &  0.5 & 2  & 0.5  & 3.9 &380& 0.1&4  \\ 
\hline
\end{tabular}
\caption{Parameters for the reference N-body models of M31 and NGC 205. The effective surface density and radius for the bulge are $\Sigma_e$ and $R_e$ respectively. The \ser index is $n_s$.}
\protect\label{tab:sbpar}
\end{center}
\end{table*}

\begin{table*}
\begin{center}
\begin{tabular}{|cc|cccc|}
\hline
Simulation& Number & $V_Y$ [$\kms$] & $M_{\rm NGC 205}$ \%& $M_{\rm M31}$  \%& $r_{-2}$  \%\\
\hline
sim25	&1 & 25 & 100 & 100 & 100 \\
sim30	&2 & 30 & 100 & 100 & 100 \\
sim35	&3 & 35 & 100 & 100 & 100 \\
sim40	&4 & 40 & 100 & 100 & 100 \\
sim50	&5 & 50 & 100 & 100 & 100 \\
sim60	&6 & 60 & 100 & 100 & 100 \\
sim80	&7 & 80 & 100 & 100 & 100 \\
sim30m03	&8 & 30 & 30 & 100 & 100 \\
sim30	&9 & 30 & 100 & 100 & 100 \\
sim30m15	&10 & 30 & 150 & 100 & 100 \\
sim30a08	&11 & 30 & 100 & 80 & 100 \\
sim30	&12 & 30 & 100 & 100 & 100 \\
sim30a15	&13& 30 & 100 & 150 & 100 \\
sim30r08	&14 & 30 & 100 & 100 & 80 \\
sim30	&15 & 30 & 100 & 100 & 100 \\
sim30r12	&16 & 30 & 100 & 100 & 120 \\
sim50r08	&17 & 50 & 100 & 100 & 80 \\
sim50	&18 & 50 & 100 & 100 & 100 \\
sim50r12	&19 & 50 & 100 & 100 & 120 \\
sim60r08	&20 & 60 & 100 & 100 & 80 \\
sim60	&21 & 60 & 100 & 100 & 100 \\
sim60r12	&22 & 60 & 100 & 100 & 120 \\
sim30a125r12	& 23 & 30 & 100 & 125 & 120 \\
sim35a125r10	&24 & 35 & 100 & 125 & 100 \\
sim35a125r12	&25 & 35 & 100 & 125 & 120 \\
sim35a125r15	&26 & 35 & 100 & 125 & 150 \\
sim35a125r20	&27 & 35 & 100 & 125 & 200 \\
sim50a125r15	&28 & 50 & 100 & 125 & 150 \\
sim65a125r15	&29 & 65 & 100 & 125 & 150 \\

\hline
\end{tabular}
\caption{Designations and parameters for the series of simulations. The second column gives the number of the simulation which corresponds to the $x$ axis values in Figs~\ref{fig:scen2}-\ref{fig:scen4} and Fig~\ref{fig:epssim}. The third column gives the initial tangential velocity and the following three columns are the scalings of the dark matter halo masses of NGC 205 and M31 and the radial scaling of the sub-subhalo distribution relative to the reference parameters from table~\ref{tab:sbpar}. A value of 100\% means the parameter is identical to the reference parameter.}
\protect\label{tab:sims}
\end{center}
\end{table*}

\section{acknowledgements} The authors thank the referee and the editor for their constructive comments. GWA is a postdoctoral fellow of the FWO Vlaanderen (Belgium). AD acknowledges the INFN grant InDark, and  the grant PRIN 2012 ``Fisica Astroparticellare Teorica" of the Italian Ministry of University and Research. The authors are grateful for the HPC resources provided by the VUB/ULB's Hydra cluster.

\appendix
\section{Supplementary figures}

Here we present supplementary figures that are referenced in the main text.
\begin{figure*}
\includegraphics[width=17.0cm]{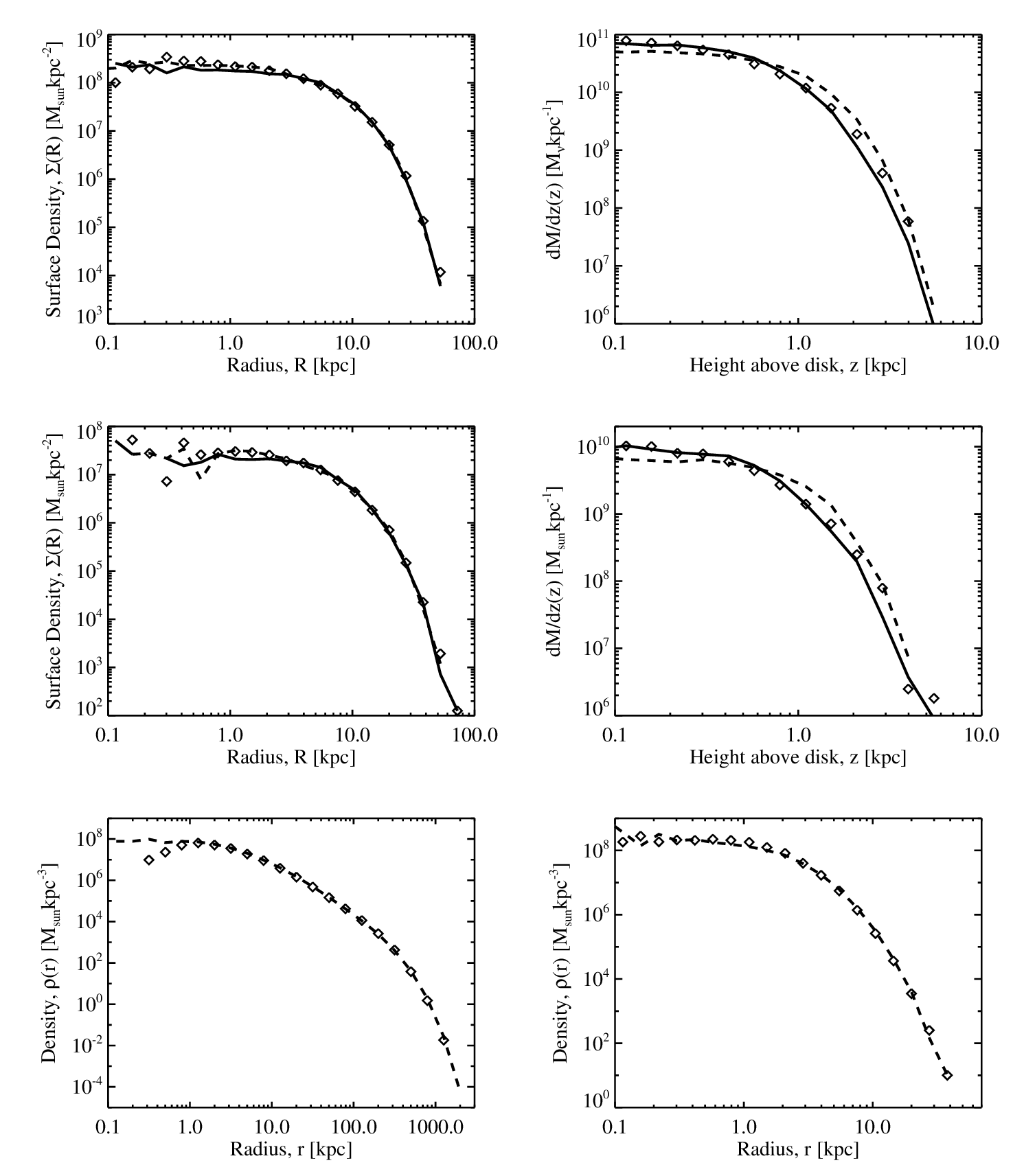}
\caption{Density profile evolution over 10~Gyr for the four mass components of M31. The top two panels show the stellar disk, where the left hand panel is the face-on surface density as a function of radius and the right hand panel is the edge-on vertical mass gradient. The symbols show the initial, analytic density and the dashed lines show the density after 10~Gyr. Some panels have solid black lines which show the density after 1.5~Gyr. The second row shows the same, but for the gaseous disk. The bottom left panel is the dark matter density and the bottom right is the bulge density.}
\label{fig:evo1}
\end{figure*}

\begin{figure*}
\includegraphics[width=17.0cm]{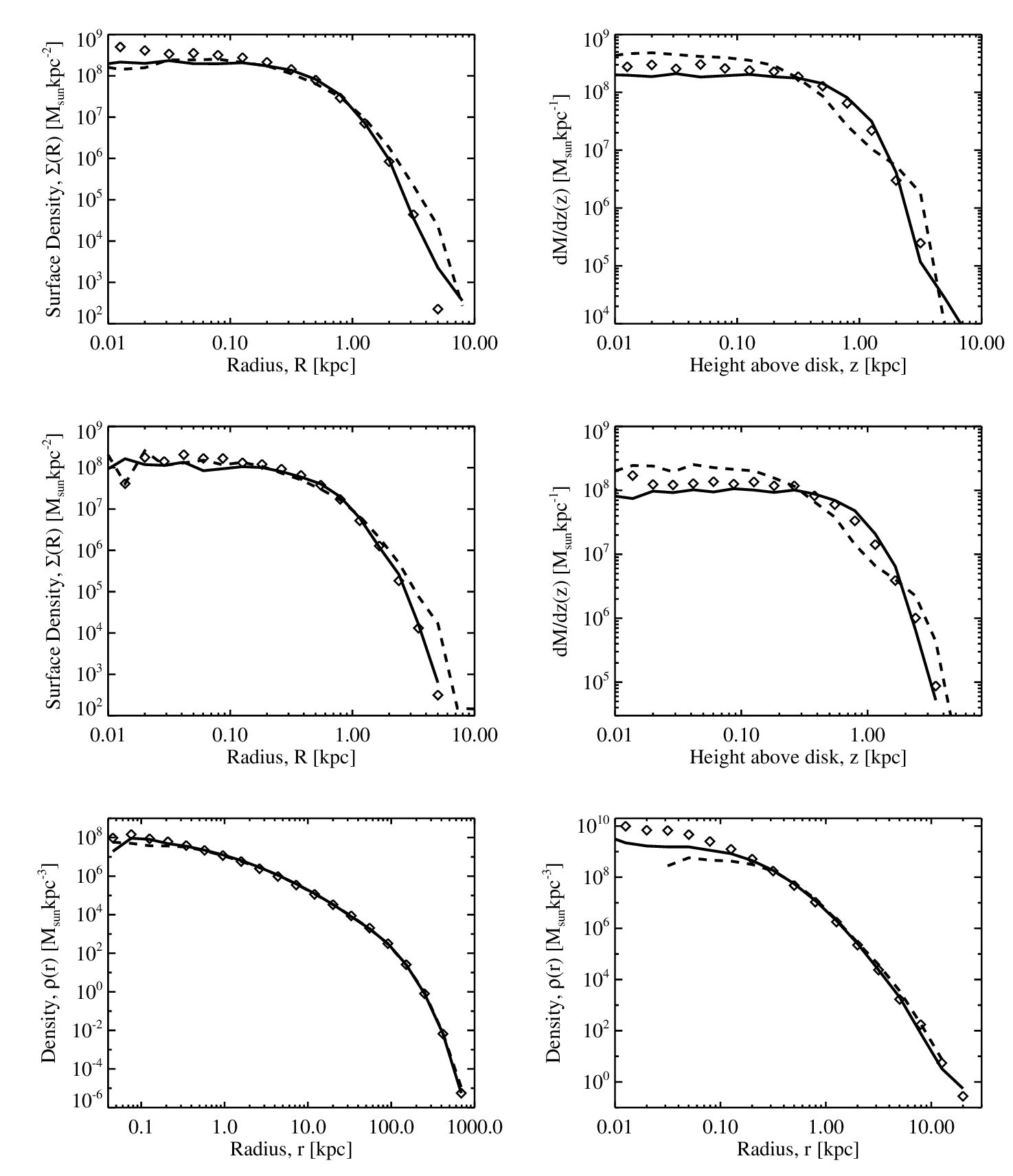}
\caption{Density profile evolution over 10 Gyr for the four mass components of NGC 205. Panels, lines and symbols as per Fig~\ref{fig:evo1}.}
\label{fig:evo2}
\end{figure*}

\begin{figure}
\includegraphics[width=8.50cm]{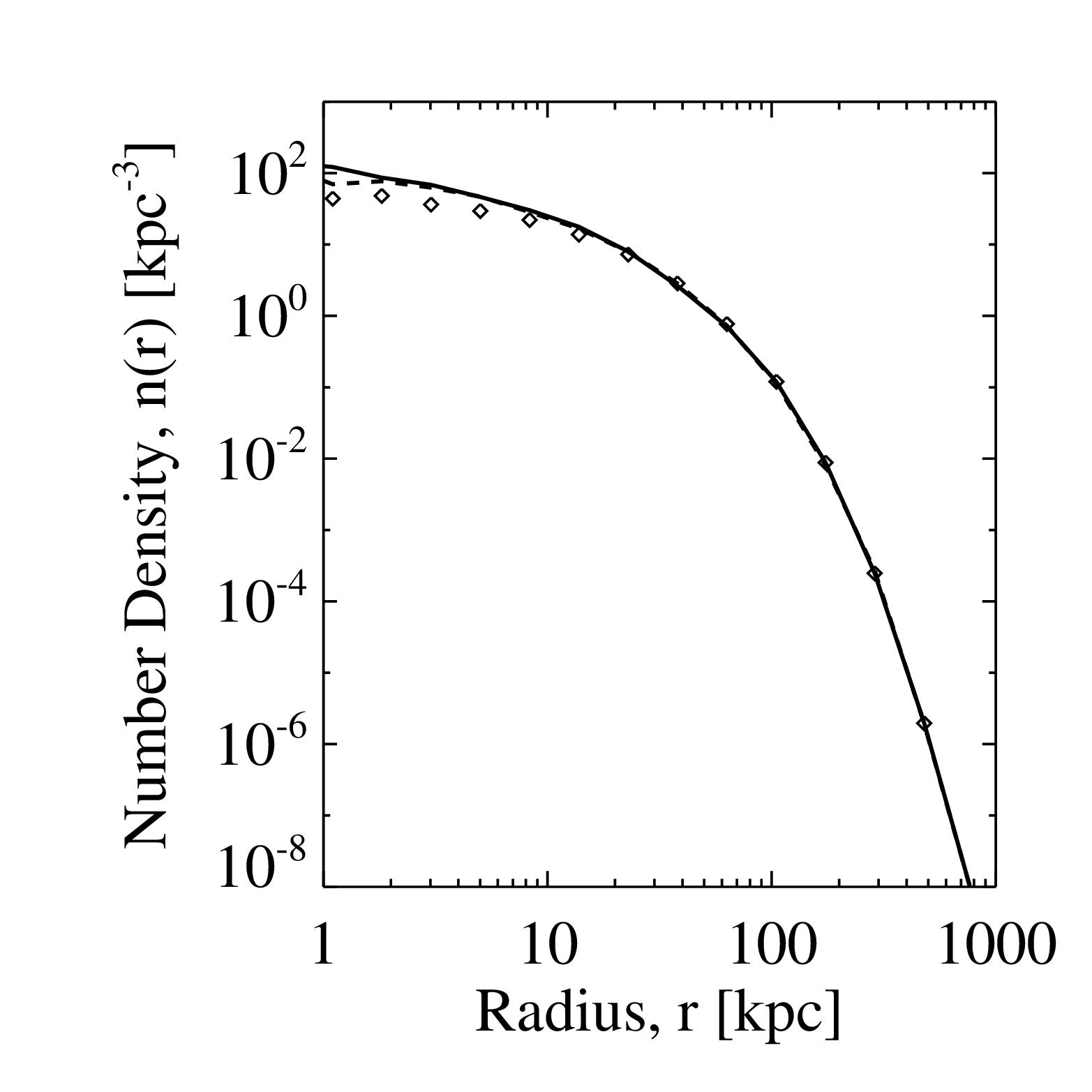}
\caption{Number density evolution over 10~Gyr of the sub-subhalo distribution surrounding NGC~205. Lines and symbols as per Fig~\ref{fig:evo1}.}
\label{fig:evo3}
\end{figure}

\begin{figure}
\includegraphics[width=8.5cm]{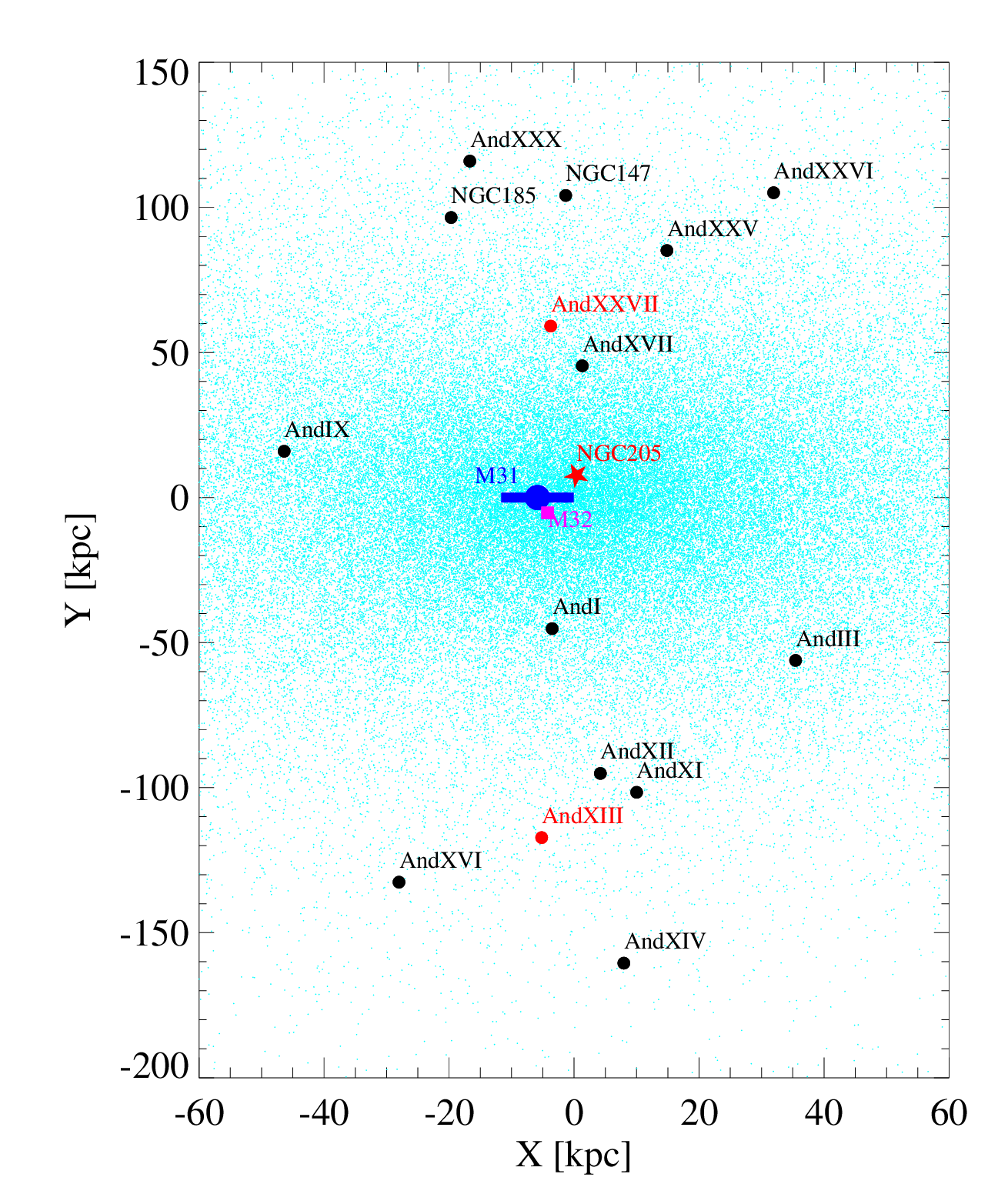}
\caption{As per Fig~\ref{fig:xya}, but for the initial distribution of sub-subhalos.}
\label{fig:xyb}
\end{figure}

\begin{figure}
\includegraphics[width=8.5cm]{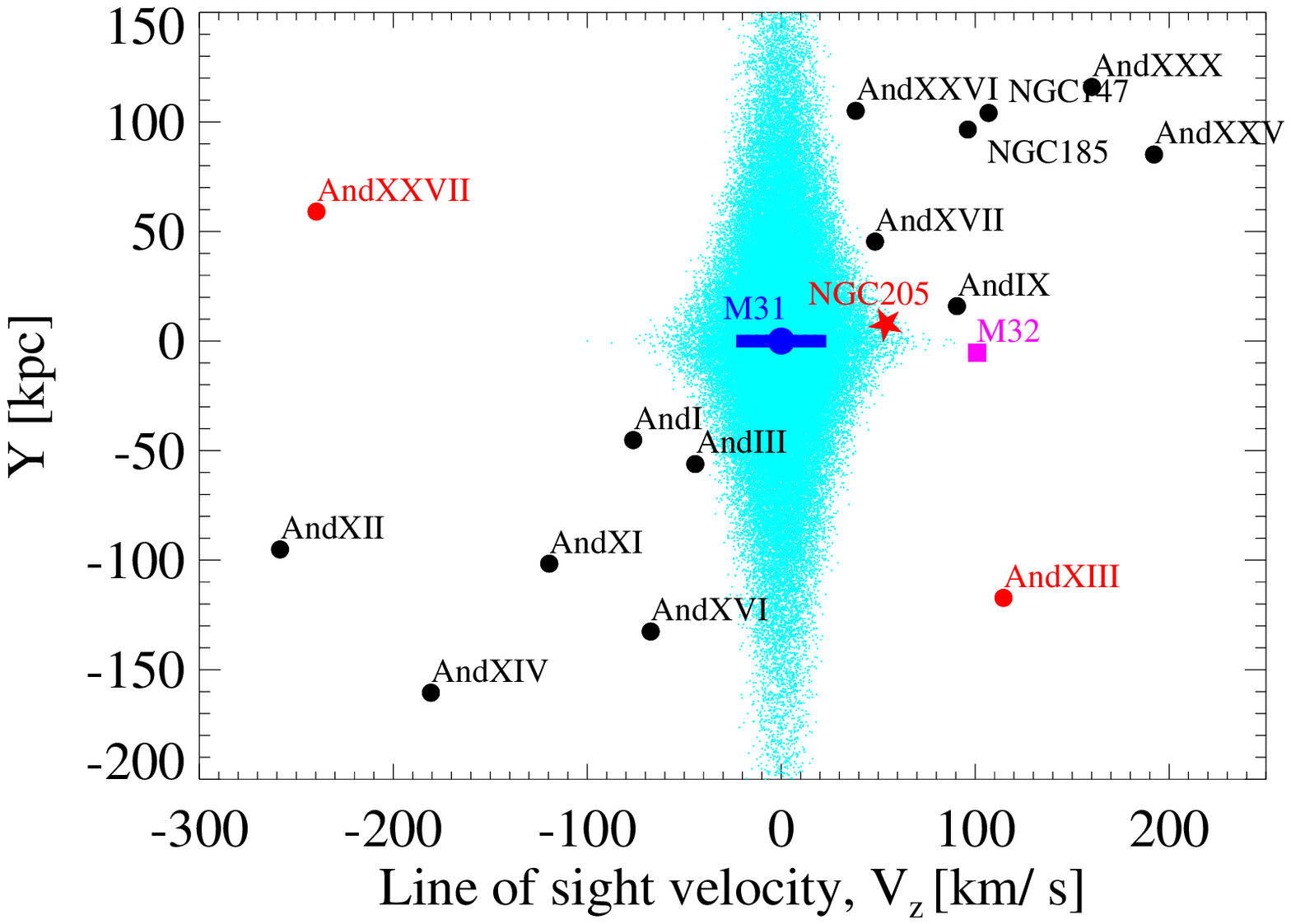}
\caption{As per Fig~\ref{fig:vzy}, but for the initial distribution of sub-subhalos.}
\label{fig:vzyb}
\end{figure}

\subsection{Simple $\chi^2$-like for the 2D comparisons}
\protect\label{sec:eps}

In addition to the Monte Carlo sampling performed in \S\ref{sec:pmc2}, we sought a simpler method to gauge which simulation snapshot, or set of initial conditions, best matches the observed $V_Z$-$Y$ and $Z$-$Y$ positions of the satellites. We began by finding the continuous series of points in both planes which are the coordinates close to the centre of the turquoise band of sub-subhalos. These were selected by finding the highest density of points along the $x$-axis (either the $V_Z$ or $Z$ coordinate) at every 5~kpc in the $Y$ coordinate (see the black lines in Figs~\ref{fig:vzy} \& \ref{fig:zy}). For each satellite position ($V_{z; \rm sat}^j$, $Y_{\rm sat}^j$), there is a coordinate ($V_{z_p}^j$, $Y_p^j$), which is a point on the black line of Fig~\ref{fig:vzy} (or \ref{fig:zy} for the $Z$-$Y$ plane), that minimises the ``distance", $\lambda_{\rm V_Z, Y}=\sqrt{({\Delta V_Z \over 1~\kms})^2+({\Delta Y \over 1 \rm kpc})^2}$ to the satellite. Of course, the weighting of the two coordinates, $V_Z$ and $Y$, is important, but since $Y$ spans roughly 300~kpc and $V_Z$ spans roughly $500~\kms$ we simply chose to weight by 1~kpc and 1~$\kms$. Crucially, the path of the turquoise band changes depending on the initial conditions. 

To give a rough estimate of which set of initial conditions and which orbital scenario (1 or 2) gives the better match to the observed satellites in the $V_Z$-$Y$ plane, we computed a figure of merit $\epsilon_{V_Z,Y}$ which is calculated by taking each observed satellite in turn and counting how many sub-subhalos, $n_i$, are in a thin bar of 10~kpc thickness between it and the point ($V_{z_p}^j$, $Y_p^j$) - that was determined above using the minimum distance criterion - on the black line in Fig~\ref{fig:vzy}, and how many are beyond it, called $n_o$. We use this to find the $\chi^2$ of that point from the inverse error function of $\sqrt{2}\rm erf^{-1}({n_o \over n_o + n_i})$. We square and sum this quantity over all 13  satellites and divide by 13 and take the square root. This should give an indication of the  probability of finding a satellite at a certain separation from the turquoise band.

\begin{figure}
\includegraphics[width=8.5cm]{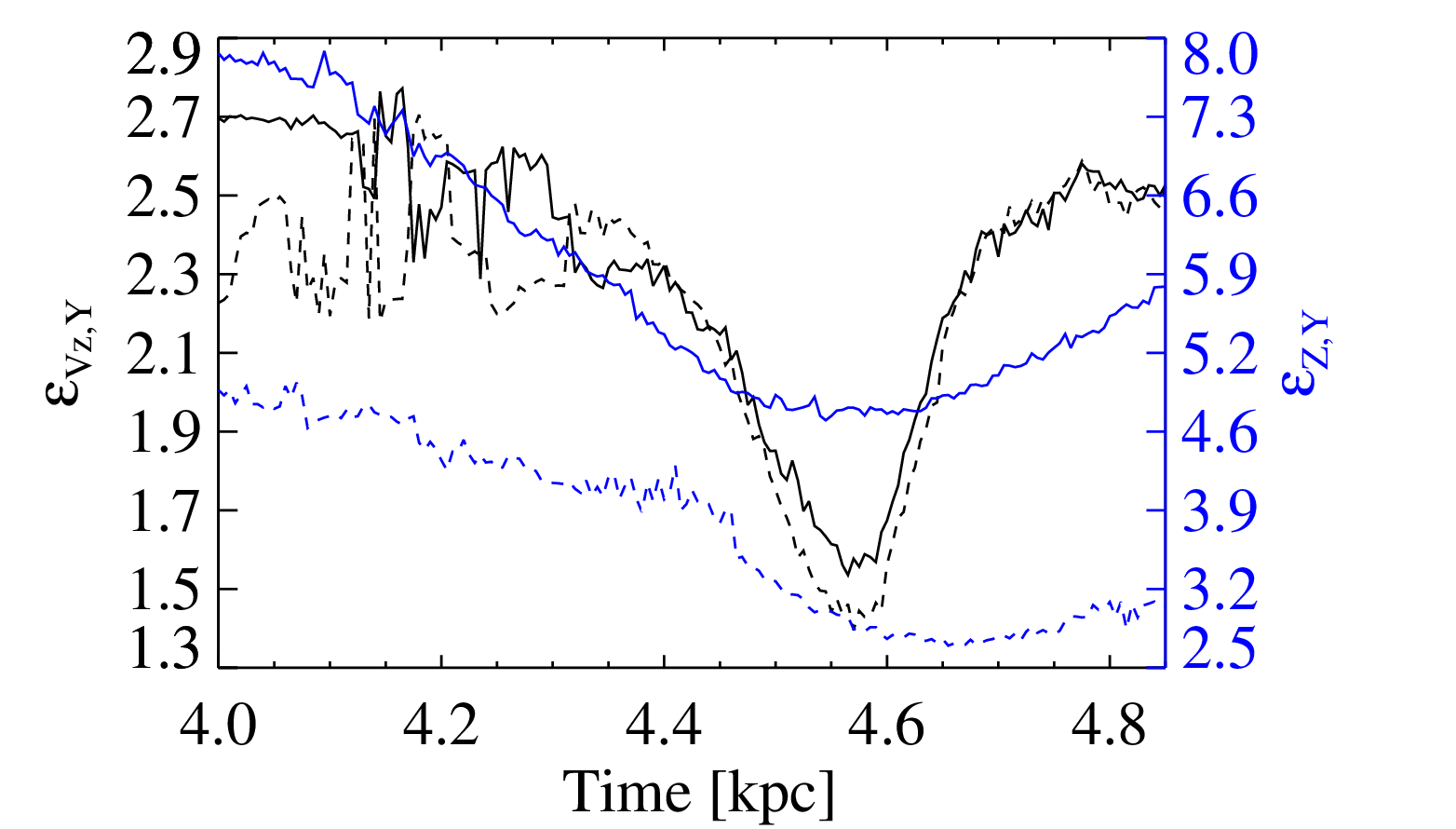}
\caption{The black lines, which correspond to the left hand $y$-axis are the $\chi^2$-like figures of merit discussed in \S\ref{sec:eps} for the match between the observed satellites and simulated sub-subhalos in the $V_Z$-$Y$ plane. The blue lines, which correspond to the right hand $y$-axis are similar figures of merit for the $Z$-$Y$ plane. The solid lines refer to scenario 1 and the dashed lines to scenario 2.}
\label{fig:eps}
\end{figure}

In Fig~\ref{fig:eps} we plot the variation of $\epsilon_{V_Z, Y}$ for the reference simulation with time near pericentre with black lines that correspond to the left hand $y$-axis. We do this for both scenario 1 (solid line) and 2 (dashed line). For both scenarios there is a factor of two improvement at pericentre over the original value.

We computed another figure of merit which was again based on the mean path of the simulated sub-subhalos in the $Z$-$Y$ plane (the black line in Fig~\ref{fig:zy}). We call this $\epsilon_{Z,Y}$, but adopted a different approach because of the larger potential uncertainty in the $Z$ measurements for each satellite. First we find the point ($Z_p^j$, $Y_p^j$) on the black line which has the same $Y$ coordinate as the satellite position ($Z_{\rm sat}^j$, $Y_{\rm sat}^j$) and then we compute the variance of $Z_p$ from $Z_i$ based on the uncertainty of $Z_i$ i.e. $({Z_p^j-Z_{\rm sat}^j \over \sigma_{Z_{\rm sat}^j}})^2$. We sum these values for all \numb  satellites, then divide by \numb and take the square root. We also plot the variation of $\epsilon_{Z,Y}$ with time in Fig~\ref{fig:eps} with blue lines corresponding to the right hand $y$-axis. Scenario 2 (dashed line) seems to be slightly more consistent with the distribution of the satellites in this plane.

\begin{figure*}
\includegraphics[width=17.0cm]{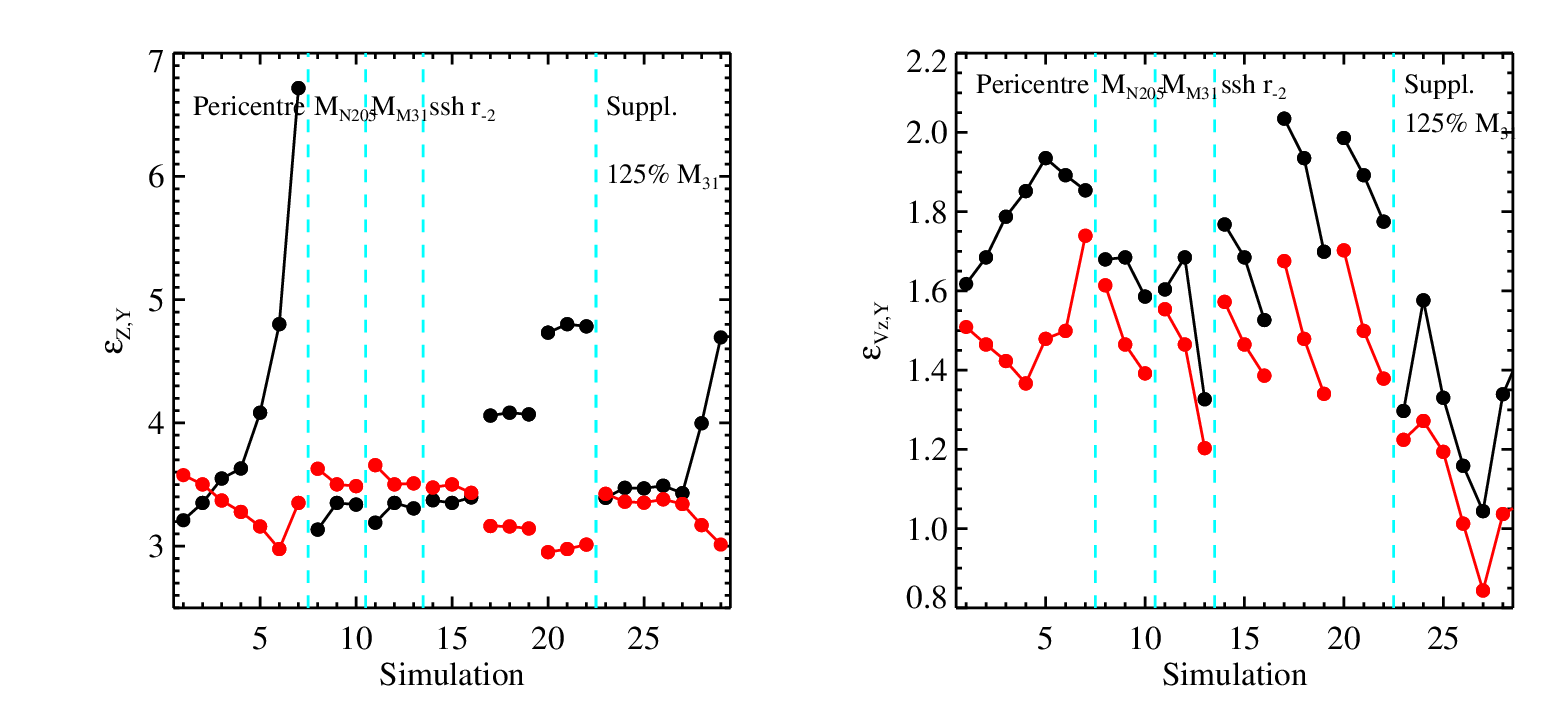}
\caption{These two panels show the variations, across our series of simulations listed in table~\ref{tab:sims}, in our $\chi^2$-like parameters found near pericentre for the distribution of simulated sub-subhalos relative to the observed satellites in the $Z$-$Y$ plane (left hand panel) and the $V_Z$-$Y$ plane (right hand panel) as initially discussed in \S\ref{sec:eps} and further discussed in \S\ref{sec:ks2}. The black lines and symbols refer to scenario 1 and the red to scenario 2. The turquoise lines isolate related simulations to show the influence of (typically) a single parameter. }
\label{fig:epssim}
\end{figure*}

For the two panels of Fig~\ref{fig:epssim} which show $\epsilon_{Z,Y}$ and $\epsilon_{V_Z,Y}$ we use black for scenario 1 and red for scenario 2. There is little improvement from scenario 1 to 2 as long as the initial tangential velocity is less than $50~\kms$. An initial tangential velocity of $50~\kms$ or greater for scenario 1 is disfavoured. From the right hand panel there is a slight preference for scenario 2, especially for larger pericentres. However, for scenario 1, most pericentres are consistent with even the most stringent distance uncertainties for NGC 205. For scenario 2, the larger the pericentre the more tension there is with the measured distance and its reliability. 

\begin{thebibliography}{111}
\expandafter\ifx\csname natexlab\endcsname\relax\def\natexlab#1{#1}\fi

\bibitem[{{Angus} {et~al.}(2011){Angus}, {Diaferio}, \& {Kroupa}}]{adk11}
{Angus} G.~W., {Diaferio} A., {Kroupa} P., 2011, \mnras, 416, 1401

\bibitem[{{Angus} {et~al.}(2015){Angus}, {Gentile}, {Swaters}, {Famaey},
  {Diaferio}, {McGaugh}, \& {Heyden}}]{angus15}
{Angus} G.~W., {Gentile} G., {Swaters} R., {Famaey} B., {Diaferio} A.,
  {McGaugh} S.~S., {Heyden} K.~J.~v.~d., 2015, \mnras, 451, 3551

\bibitem[{{Baes} \& {Dejonghe}(2002)}]{baes02}
{Baes} M., {Dejonghe} H., 2002, \aap, 393, 485

\bibitem[{{Bahl} \& {Baumgardt}(2014)}]{bahl14}
{Bahl} H., {Baumgardt} H., 2014, \mnras, 438, 2916

\bibitem[{{Bell} \& {de Jong}(2001)}]{belldejong}
{Bell} E.~F., {de Jong} R.~S., 2001, \apj, 550, 212

\bibitem[{{Benson} {et~al.}(2002){Benson}, {Frenk}, {Lacey}, {Baugh}, \&
  {Cole}}]{benson02}
{Benson} A.~J., {Frenk} C.~S., {Lacey} C.~G., {Baugh} C.~M., {Cole} S., 2002,
  \mnras, 333, 177

\bibitem[{{Bershady} {et~al.}(2010){Bershady}, {Verheijen}, {Westfall},
  {Andersen}, {Swaters}, \& {Martinsson}}]{bershady10b}
{Bershady} M.~A., {Verheijen} M.~A.~W., {Westfall} K.~B., {Andersen} D.~R.,
  {Swaters} R.~A., {Martinsson} T., 2010, \apj, 716, 234

\bibitem[{{Boylan-Kolchin} {et~al.}(2014){Boylan-Kolchin}, {Bullock}, \&
  {Garrison-Kimmel}}]{boylan14}
{Boylan-Kolchin} M., {Bullock} J.~S., {Garrison-Kimmel} S., 2014, \mnras, 443,
  L44

\bibitem[{{Boylan-Kolchin} {et~al.}(2011){Boylan-Kolchin}, {Bullock}, \&
  {Kaplinghat}}]{boylan11}
{Boylan-Kolchin} M., {Bullock} J.~S., {Kaplinghat} M., 2011, \mnras, 415, L40

\bibitem[{{Boylan-Kolchin} {et~al.}(2012){Boylan-Kolchin}, {Bullock}, \&
  {Kaplinghat}}]{boylan12}
---, 2012, \mnras, 422, 1203

\bibitem[{{Bullock} {et~al.}(2000){Bullock}, {Kravtsov}, \&
  {Weinberg}}]{bullock00}
{Bullock} J.~S., {Kravtsov} A.~V., {Weinberg} D.~H., 2000, \apj, 539, 517

\bibitem[{{Carignan} {et~al.}(2006){Carignan}, {Chemin}, {Huchtmeier}, \&
  {Lockman}}]{carignan06}
{Carignan} C., {Chemin} L., {Huchtmeier} W.~K., {Lockman} F.~J., 2006, \apjl,
  641, L109

\bibitem[{{Castellano} {et~al.}(2016){Castellano}, {Dayal}, {Pentericci},
  {Fontana}, {Hutter}, {Brammer}, {Merlin}, {Grazian}, {Pilo}, {Amorin},
  {Cristiani}, {Dickinson}, {Ferrara}, {Gallerani}, {Giallongo}, {Giavalisco},
  {Guaita}, {Koekemoer}, {Maiolino}, {Paris}, {Santini}, {Vallini}, {Vanzella},
  \& {Wagg}}]{castellano16}
{Castellano} M., {Dayal} P., {Pentericci} L., {Fontana} A., {Hutter} A.,
  {Brammer} G., {Merlin} E., {Grazian} A., {Pilo} S., {Amorin} R., {Cristiani}
  S., {Dickinson} M., {Ferrara} A., {Gallerani} S., {Giallongo} E.,
  {Giavalisco} M., {Guaita} L., {Koekemoer} A., {Maiolino} R., {Paris} D.,
  {Santini} P., {Vallini} L., {Vanzella} E., {Wagg} J., 2016, \apjl, 818, L3

\bibitem[{{Cautun} {et~al.}(2015{\natexlab{a}}){Cautun}, {Bose}, {Frenk},
  {Guo}, {Han}, {Hellwing}, {Sawala}, \& {Wang}}]{cautun15b}
{Cautun} M., {Bose} S., {Frenk} C.~S., {Guo} Q., {Han} J., {Hellwing} W.~A.,
  {Sawala} T., {Wang} W., 2015{\natexlab{a}}, \mnras, 452, 3838

\bibitem[{{Cautun} {et~al.}(2015{\natexlab{b}}){Cautun}, {Wang}, {Frenk}, \&
  {Sawala}}]{cautun15a}
{Cautun} M., {Wang} W., {Frenk} C.~S., {Sawala} T., 2015{\natexlab{b}}, \mnras,
  449, 2576

\bibitem[{{Conn} {et~al.}(2012){Conn}, {Ibata}, {Lewis}, {Parker}, {Zucker},
  {Martin}, {McConnachie}, {Irwin}, {Tanvir}, {Fardal}, {Ferguson}, {Chapman},
  \& {Valls-Gabaud}}]{conn2}
{Conn} A.~R., {Ibata} R.~A., {Lewis} G.~F., {Parker} Q.~A., {Zucker} D.~B.,
  {Martin} N.~F., {McConnachie} A.~W., {Irwin} M.~J., {Tanvir} N., {Fardal}
  M.~A., {Ferguson} A.~M.~N., {Chapman} S.~C., {Valls-Gabaud} D., 2012, \apj,
  758, 11

\bibitem[{{Conn} {et~al.}(2011){Conn}, {Lewis}, {Ibata}, {Parker}, {Zucker},
  {McConnachie}, {Martin}, {Irwin}, {Tanvir}, {Fardal}, \& {Ferguson}}]{conn1}
{Conn} A.~R., {Lewis} G.~F., {Ibata} R.~A., {Parker} Q.~A., {Zucker} D.~B.,
  {McConnachie} A.~W., {Martin} N.~F., {Irwin} M.~J., {Tanvir} N., {Fardal}
  M.~A., {Ferguson} A.~M.~N., 2011, \apj, 740, 69

\bibitem[{{Conn} {et~al.}(2013){Conn}, {Lewis}, {Ibata}, {Parker}, {Zucker},
  {McConnachie}, {Martin}, {Valls-Gabaud}, {Tanvir}, {Irwin}, {Ferguson}, \&
  {Chapman}}]{conn13}
{Conn} A.~R., {Lewis} G.~F., {Ibata} R.~A., {Parker} Q.~A., {Zucker} D.~B.,
  {McConnachie} A.~W., {Martin} N.~F., {Valls-Gabaud} D., {Tanvir} N., {Irwin}
  M.~J., {Ferguson} A.~M.~N., {Chapman} S.~C., 2013, \apj, 766, 120

\bibitem[{{Corbelli} {et~al.}(2010){Corbelli}, {Lorenzoni}, {Walterbos},
  {Braun}, \& {Thilker}}]{corbelli10}
{Corbelli} E., {Lorenzoni} S., {Walterbos} R., {Braun} R., {Thilker} D., 2010,
  \aap, 511, A89

\bibitem[{{Crnojevi{\'c}} {et~al.}(2014){Crnojevi{\'c}}, {Ferguson}, {Irwin},
  {McConnachie}, {Bernard}, {Fardal}, {Ibata}, {Lewis}, {Martin}, {Navarro},
  {No{\"e}l}, \& {Pasetto}}]{crno}
{Crnojevi{\'c}} D., {Ferguson} A.~M.~N., {Irwin} M.~J., {McConnachie} A.~W.,
  {Bernard} E.~J., {Fardal} M.~A., {Ibata} R.~A., {Lewis} G.~F., {Martin}
  N.~F., {Navarro} J.~F., {No{\"e}l} N.~E.~D., {Pasetto} S., 2014, \mnras, 445,
  3862

\bibitem[{{Davis} {et~al.}(1985){Davis}, {Efstathiou}, {Frenk}, \&
  {White}}]{davis85}
{Davis} M., {Efstathiou} G., {Frenk} C.~S., {White} S.~D.~M., 1985, \apj, 292,
  371

\bibitem[{{de Grijs} \& {Bono}(2014)}]{degrijs14}
{de Grijs} R., {Bono} G., 2014, \aj, 148, 17

\bibitem[{{De Rijcke} {et~al.}(2006){De Rijcke}, {Prugniel}, {Simien}, \&
  {Dejonghe}}]{derijcke06}
{De Rijcke} S., {Prugniel} P., {Simien} F., {Dejonghe} H., 2006, \mnras, 369,
  1321

\bibitem[{{Deason} {et~al.}(2015){Deason}, {Wetzel}, {Garrison-Kimmel}, \&
  {Belokurov}}]{deason15}
{Deason} A.~J., {Wetzel} A.~R., {Garrison-Kimmel} S., {Belokurov} V., 2015,
  \mnras, 453, 3568

\bibitem[{{Diemand} {et~al.}(2007){Diemand}, {Kuhlen}, \& {Madau}}]{diemand07}
{Diemand} J., {Kuhlen} M., {Madau} P., 2007, \apj, 657, 262

\bibitem[{{Diemand} {et~al.}(2004){Diemand}, {Moore}, \& {Stadel}}]{diemand04}
{Diemand} J., {Moore} B., {Stadel} J., 2004, \mnras, 352, 535

\bibitem[{{Dierickx} {et~al.}(2014){Dierickx}, {Blecha}, \&
  {Loeb}}]{dierickx14}
{Dierickx} M., {Blecha} L., {Loeb} A., 2014, \apjl, 788, L38

\bibitem[{{Dijkstra} {et~al.}(2004){Dijkstra}, {Haiman}, {Rees}, \&
  {Weinberg}}]{dijkstra04}
{Dijkstra} M., {Haiman} Z., {Rees} M.~J., {Weinberg} D.~H., 2004, \apj, 601,
  666

\bibitem[{{D'Onghia} \& {Lake}(2008)}]{donghia08}
{D'Onghia} E., {Lake} G., 2008, \apjl, 686, L61

\bibitem[{{Efstathiou}(1992)}]{efstathiou92}
{Efstathiou} G., 1992, \mnras, 256, 43P

\bibitem[{{Evslin}(2014)}]{evslin14}
{Evslin} J., 2014, \mnras, 440, 1225

\bibitem[{{Fouquet} {et~al.}(2012){Fouquet}, {Hammer}, {Yang}, {Puech}, \&
  {Flores}}]{fouquet12}
{Fouquet} S., {Hammer} F., {Yang} Y., {Puech} M., {Flores} H., 2012, \mnras,
  427, 1769

\bibitem[{{Frenk} {et~al.}(1988){Frenk}, {White}, {Davis}, \&
  {Efstathiou}}]{frenk88}
{Frenk} C.~S., {White} S.~D.~M., {Davis} M., {Efstathiou} G., 1988, \apj, 327,
  507

\bibitem[{{Frenk} {et~al.}(1985){Frenk}, {White}, {Efstathiou}, \&
  {Davis}}]{frenk85}
{Frenk} C.~S., {White} S.~D.~M., {Efstathiou} G., {Davis} M., 1985, \nat, 317,
  595

\bibitem[{{Garrison-Kimmel} {et~al.}(2014{\natexlab{a}}){Garrison-Kimmel},
  {Boylan-Kolchin}, {Bullock}, \& {Kirby}}]{garrison14}
{Garrison-Kimmel} S., {Boylan-Kolchin} M., {Bullock} J.~S., {Kirby} E.~N.,
  2014{\natexlab{a}}, \mnras, 444, 222

\bibitem[{{Garrison-Kimmel} {et~al.}(2014{\natexlab{b}}){Garrison-Kimmel},
  {Boylan-Kolchin}, {Bullock}, \& {Lee}}]{elvis}
{Garrison-Kimmel} S., {Boylan-Kolchin} M., {Bullock} J.~S., {Lee} K.,
  2014{\natexlab{b}}, \mnras, 438, 2578

\bibitem[{{Garrison-Kimmel} {et~al.}(2013){Garrison-Kimmel}, {Rocha},
  {Boylan-Kolchin}, {Bullock}, \& {Lally}}]{garrison13}
{Garrison-Kimmel} S., {Rocha} M., {Boylan-Kolchin} M., {Bullock} J.~S., {Lally}
  J., 2013, \mnras, 433, 3539

\bibitem[{{Geha} {et~al.}(2006){Geha}, {Guhathakurta}, {Rich}, \&
  {Cooper}}]{geha06}
{Geha} M., {Guhathakurta} P., {Rich} R.~M., {Cooper} M.~C., 2006, \aj, 131, 332

\bibitem[{{Gentile} {et~al.}(2007){Gentile}, {Famaey}, {Combes}, {Kroupa},
  {Zhao}, \& {Tiret}}]{gentile07a}
{Gentile} G., {Famaey} B., {Combes} F., {Kroupa} P., {Zhao} H.~S., {Tiret} O.,
  2007, \aap, 472, L25

\bibitem[{{Graham} \& {Driver}(2005)}]{graham05}
{Graham} A.~W., {Driver} S.~P., 2005, PASA, 22, 118

\bibitem[{{Griffen} {et~al.}(2016){Griffen}, {Ji}, {Dooley}, {G{\'o}mez},
  {Vogelsberger}, {O'Shea}, \& {Frebel}}]{griffen16}
{Griffen} B.~F., {Ji} A.~P., {Dooley} G.~A., {G{\'o}mez} F.~A., {Vogelsberger}
  M., {O'Shea} B.~W., {Frebel} A., 2016, \apj, 818, 10

\bibitem[{{Hammer} {et~al.}(2013){Hammer}, {Yang}, {Fouquet}, {Pawlowski},
  {Kroupa}, {Puech}, {Flores}, \& {Wang}}]{hammer13}
{Hammer} F., {Yang} Y., {Fouquet} S., {Pawlowski} M.~S., {Kroupa} P., {Puech}
  M., {Flores} H., {Wang} J., 2013, \mnras, 431, 3543

\bibitem[{{Hammer} {et~al.}(2015){Hammer}, {Yang}, {Flores}, {Puech}, \&
  {Fouquet}}]{hammer15}
{Hammer} F., {Yang} Y.~B., {Flores} H., {Puech} M., {Fouquet} S., 2015, \apj,
  813, 110

\bibitem[{{Hernquist}(1990)}]{hernquist90}
{Hernquist} L., 1990, \apj, 356, 359

\bibitem[{{Hernquist}(1993)}]{hernquist93}
---, 1993, \apjs, 86, 389

\bibitem[{{Howley} {et~al.}(2008){Howley}, {Geha}, {Guhathakurta},
  {Montgomery}, {Laughlin}, \& {Johnston}}]{howley08}
{Howley} K.~M., {Geha} M., {Guhathakurta} P., {Montgomery} R.~M., {Laughlin}
  G., {Johnston} K.~V., 2008, \apj, 683, 722

\bibitem[{{Ibata} {et~al.}(2014{\natexlab{a}}){Ibata}, {Ibata}, {Famaey}, \&
  {Lewis}}]{ibata14}
{Ibata} N.~G., {Ibata} R.~A., {Famaey} B., {Lewis} G.~F., 2014{\natexlab{a}},
  \nat, 511, 563

\bibitem[{{Ibata} {et~al.}(2015){Ibata}, {Famaey}, {Lewis}, {Ibata}, \&
  {Martin}}]{ibata15}
{Ibata} R.~A., {Famaey} B., {Lewis} G.~F., {Ibata} N.~G., {Martin} N., 2015,
  \apj, 805, 67

\bibitem[{{Ibata} {et~al.}(2014{\natexlab{b}}){Ibata}, {Ibata}, {Lewis},
  {Martin}, {Conn}, {Elahi}, {Arias}, \& {Fernando}}]{ibata14a}
{Ibata} R.~A., {Ibata} N.~G., {Lewis} G.~F., {Martin} N.~F., {Conn} A., {Elahi}
  P., {Arias} V., {Fernando} N., 2014{\natexlab{b}}, \apjl, 784, L6

\bibitem[{{Ibata} {et~al.}(2013){Ibata}, {Lewis}, {Conn}, {Irwin},
  {McConnachie}, {Chapman}, {Collins}, {Fardal}, {Ferguson}, {Ibata}, {Mackey},
  {Martin}, {Navarro}, {Rich}, {Valls-Gabaud}, \& {Widrow}}]{ibata13}
{Ibata} R.~A., {Lewis} G.~F., {Conn} A.~R., {Irwin} M.~J., {McConnachie} A.~W.,
  {Chapman} S.~C., {Collins} M.~L., {Fardal} M., {Ferguson} A.~M.~N., {Ibata}
  N.~G., {Mackey} A.~D., {Martin} N.~F., {Navarro} J., {Rich} R.~M.,
  {Valls-Gabaud} D., {Widrow} L.~M., 2013, \nat, 493, 62

\bibitem[{{Jensen} {et~al.}(2003){Jensen}, {Tonry}, {Barris}, {Thompson},
  {Liu}, {Rieke}, {Ajhar}, \& {Blakeslee}}]{jensen03}
{Jensen} J.~B., {Tonry} J.~L., {Barris} B.~J., {Thompson} R.~I., {Liu} M.~C.,
  {Rieke} M.~J., {Ajhar} E.~A., {Blakeslee} J.~P., 2003, \apj, 583, 712

\bibitem[{{Kallivayalil} {et~al.}(2006{\natexlab{a}}){Kallivayalil}, {van der
  Marel}, \& {Alcock}}]{kallivayalil06b}
{Kallivayalil} N., {van der Marel} R.~P., {Alcock} C., 2006{\natexlab{a}},
  \apj, 652, 1213

\bibitem[{{Kallivayalil} {et~al.}(2006{\natexlab{b}}){Kallivayalil}, {van der
  Marel}, {Alcock}, {Axelrod}, {Cook}, {Drake}, \& {Geha}}]{kallivayalil06a}
{Kallivayalil} N., {van der Marel} R.~P., {Alcock} C., {Axelrod} T., {Cook}
  K.~H., {Drake} A.~J., {Geha} M., 2006{\natexlab{b}}, \apj, 638, 772

\bibitem[{{Kallivayalil} {et~al.}(2013){Kallivayalil}, {van der Marel},
  {Besla}, {Anderson}, \& {Alcock}}]{kallivayalil13}
{Kallivayalil} N., {van der Marel} R.~P., {Besla} G., {Anderson} J., {Alcock}
  C., 2013, \apj, 764, 161

\bibitem[{{Karachentsev} {et~al.}(2004){Karachentsev}, {Karachentseva},
  {Huchtmeier}, \& {Makarov}}]{karachentsev04}
{Karachentsev} I.~D., {Karachentseva} V.~E., {Huchtmeier} W.~K., {Makarov}
  D.~I., 2004, \aj, 127, 2031

\bibitem[{{Kazantzidis} {et~al.}(2004){Kazantzidis}, {Magorrian}, \&
  {Moore}}]{kazantzidis04}
{Kazantzidis} S., {Magorrian} J., {Moore} B., 2004, \apj, 601, 37

\bibitem[{{Kent}(1989)}]{kent89}
{Kent} S.~M., 1989, PASP, 101, 489

\bibitem[{{Klypin} {et~al.}(1999){Klypin}, {Kravtsov}, {Valenzuela}, \&
  {Prada}}]{klypin99}
{Klypin} A., {Kravtsov} A.~V., {Valenzuela} O., {Prada} F., 1999, \apj, 522, 82

\bibitem[{{Knebe} {et~al.}(2011){Knebe}, {Libeskind}, {Knollmann},
  {Martinez-Vaquero}, {Yepes}, {Gottl{\"o}ber}, \& {Hoffman}}]{knebe11}
{Knebe} A., {Libeskind} N.~I., {Knollmann} S.~R., {Martinez-Vaquero} L.~A.,
  {Yepes} G., {Gottl{\"o}ber} S., {Hoffman} Y., 2011, \mnras, 412, 529

\bibitem[{{Koch} \& {Grebel}(2006)}]{koch06}
{Koch} A., {Grebel} E.~K., 2006, \aj, 131, 1405

\bibitem[{{Kroupa}(1997)}]{kroupa97}
{Kroupa} P., 1997, \na, 2, 139

\bibitem[{{Kroupa} {et~al.}(2010){Kroupa}, {Famaey}, {de Boer},
  {Dabringhausen}, {Pawlowski}, {Boily}, {Jerjen}, {Forbes}, {Hensler}, \&
  {Metz}}]{kroupa10}
{Kroupa} P., {Famaey} B., {de Boer} K.~S., {Dabringhausen} J., {Pawlowski}
  M.~S., {Boily} C.~M., {Jerjen} H., {Forbes} D., {Hensler} G., {Metz} M.,
  2010, \aap, 523, A32

\bibitem[{{Li} \& {Helmi}(2008)}]{lihelmi08}
{Li} Y., {Helmi} A., 2008, \mnras, 385, 1365

\bibitem[{{Libeskind} {et~al.}(2005){Libeskind}, {Frenk}, {Cole}, {Helly},
  {Jenkins}, {Navarro}, \& {Power}}]{libeskind05}
{Libeskind} N.~I., {Frenk} C.~S., {Cole} S., {Helly} J.~C., {Jenkins} A.,
  {Navarro} J.~F., {Power} C., 2005, \mnras, 363, 146

\bibitem[{{Libeskind} {et~al.}(2015){Libeskind}, {Hoffman}, {Tully},
  {Courtois}, {Pomar{\`e}de}, {Gottl{\"o}ber}, \& {Steinmetz}}]{libeskind15}
{Libeskind} N.~I., {Hoffman} Y., {Tully} R.~B., {Courtois} H.~M.,
  {Pomar{\`e}de} D., {Gottl{\"o}ber} S., {Steinmetz} M., 2015, \mnras, 452,
  1052

\bibitem[{{Lovell} {et~al.}(2011){Lovell}, {Eke}, {Frenk}, \&
  {Jenkins}}]{lovell11}
{Lovell} M.~R., {Eke} V.~R., {Frenk} C.~S., {Jenkins} A., 2011, \mnras, 413,
  3013

\bibitem[{{Lynden-Bell}(1983)}]{lb83}
{Lynden-Bell} D., 1983, in IAU Symposium, Vol. 100, Internal Kinematics and
  Dynamics of Galaxies, {Athanassoula} E., ed., pp. 89--91

\bibitem[{{Mateo}(1998)}]{mateo98}
{Mateo} M.~L., 1998, \araa, 36, 435

\bibitem[{{McAlpine} {et~al.}(2016){McAlpine}, {Helly}, {Schaller}, {Trayford},
  {Qu}, {Furlong}, {Bower}, {Crain}, {Schaye}, {Theuns}, {Dalla Vecchia},
  {Frenk}, {McCarthy}, {Jenkins}, {Rosas-Guevara}, {White}, {Baes}, {Camps}, \&
  {Lemson}}]{mcalpine15}
{McAlpine} S., {Helly} J.~C., {Schaller} M., {Trayford} J.~W., {Qu} Y.,
  {Furlong} M., {Bower} R.~G., {Crain} R.~A., {Schaye} J., {Theuns} T., {Dalla
  Vecchia} C., {Frenk} C.~S., {McCarthy} I.~G., {Jenkins} A., {Rosas-Guevara}
  Y., {White} S.~D.~M., {Baes} M., {Camps} P., {Lemson} G., 2016, Astronomy and
  Computing, 15, 72

\bibitem[{{McConnachie}(2012)}]{mcconnachie12}
{McConnachie} A.~W., 2012, \aj, 144, 4

\bibitem[{{McConnachie} {et~al.}(2005){McConnachie}, {Irwin}, {Ferguson},
  {Ibata}, {Lewis}, \& {Tanvir}}]{mcconnachie05}
{McConnachie} A.~W., {Irwin} M.~J., {Ferguson} A.~M.~N., {Ibata} R.~A., {Lewis}
  G.~F., {Tanvir} N., 2005, \mnras, 356, 979

\bibitem[{{Metz} {et~al.}(2008){Metz}, {Kroupa}, \& {Libeskind}}]{metz08}
{Metz} M., {Kroupa} P., {Libeskind} N.~I., 2008, \apj, 680, 287

\bibitem[{{Metz} {et~al.}(2009){Metz}, {Kroupa}, {Theis}, {Hensler}, \&
  {Jerjen}}]{metz09}
{Metz} M., {Kroupa} P., {Theis} C., {Hensler} G., {Jerjen} H., 2009, \apj, 697,
  269

\bibitem[{{Moore} {et~al.}(1999){Moore}, {Ghigna}, {Governato}, {Lake},
  {Quinn}, {Stadel}, \& {Tozzi}}]{moore99}
{Moore} B., {Ghigna} S., {Governato} F., {Lake} G., {Quinn} T., {Stadel} J.,
  {Tozzi} P., 1999, \apj, 524, L19

\bibitem[{{Navarro} {et~al.}(1997){Navarro}, {Frenk}, \& {White}}]{nfw97}
{Navarro} J.~F., {Frenk} C.~S., {White} S.~D.~M., 1997, \apj, 490, 493

\bibitem[{{Pawlowski}(2016)}]{pawlowski16}
{Pawlowski} M.~S., 2016, \mnras, 456, 448

\bibitem[{{Pawlowski} {et~al.}(2014){Pawlowski}, {Famaey}, {Jerjen}, {Merritt},
  {Kroupa}, {Dabringhausen}, {L{\"u}ghausen}, {Forbes}, {Hensler}, {Hammer},
  {Puech}, {Fouquet}, {Flores}, \& {Yang}}]{pawlowski14}
{Pawlowski} M.~S., {Famaey} B., {Jerjen} H., {Merritt} D., {Kroupa} P.,
  {Dabringhausen} J., {L{\"u}ghausen} F., {Forbes} D.~A., {Hensler} G.,
  {Hammer} F., {Puech} M., {Fouquet} S., {Flores} H., {Yang} Y., 2014, \mnras,
  442, 2362

\bibitem[{{Pawlowski} {et~al.}(2012){Pawlowski}, {Kroupa}, {Angus}, {de Boer},
  {Famaey}, \& {Hensler}}]{pawlowski12}
{Pawlowski} M.~S., {Kroupa} P., {Angus} G., {de Boer} K.~S., {Famaey} B.,
  {Hensler} G., 2012, \mnras, 424, 80

\bibitem[{{Pawlowski} {et~al.}(2013){Pawlowski}, {Kroupa}, \&
  {Jerjen}}]{pawlowski13}
{Pawlowski} M.~S., {Kroupa} P., {Jerjen} H., 2013, \mnras, 435, 1928

\bibitem[{{Pawlowski} {et~al.}(2015){Pawlowski}, {McGaugh}, \&
  {Jerjen}}]{pawlowski15}
{Pawlowski} M.~S., {McGaugh} S.~S., {Jerjen} H., 2015, \mnras, 453, 1047

\bibitem[{{Phillips} {et~al.}(2015){Phillips}, {Cooper}, {Bullock}, \&
  {Boylan-Kolchin}}]{phillips15}
{Phillips} J.~I., {Cooper} M.~C., {Bullock} J.~S., {Boylan-Kolchin} M., 2015,
  \mnras, 453, 3839

\bibitem[{{Piatek} {et~al.}(2007){Piatek}, {Pryor}, \& {Olszewski}}]{piatek08}
{Piatek} S., {Pryor} C., {Olszewski} E.~W., 2007, preprint(arXiv:0712.1764),
  712

\bibitem[{{Planck Collaboration} {et~al.}(2014){Planck Collaboration}, {Ade},
  {Aghanim}, {Armitage-Caplan}, {Arnaud}, {Ashdown}, {Atrio-Barandela},
  {Aumont}, {Baccigalupi}, {Banday}, \& et~al.}]{planck14}
{Planck Collaboration}, {Ade} P.~A.~R., {Aghanim} N., {Armitage-Caplan} C.,
  {Arnaud} M., {Ashdown} M., {Atrio-Barandela} F., {Aumont} J., {Baccigalupi}
  C., {Banday} A.~J., et~al., 2014, \aap, 571, A16

\bibitem[{{Sales} {et~al.}(2016){Sales}, {Navarro}, {Kallivayalil}, \&
  {Frenk}}]{sales16}
{Sales} L.~V., {Navarro} J.~F., {Kallivayalil} N., {Frenk} C.~S., 2016,
  ArXiv:1605.03574

\bibitem[{{Sand} {et~al.}(2015){Sand}, {Spekkens}, {Crnojevi{\'c}}, {Hargis},
  {Willman}, {Strader}, \& {Grillmair}}]{sand15}
{Sand} D.~J., {Spekkens} K., {Crnojevi{\'c}} D., {Hargis} J.~R., {Willman} B.,
  {Strader} J., {Grillmair} C.~J., 2015, \apjl, 812, L13

\bibitem[{{Sawa} \& {Fujimoto}(2005)}]{sawa05}
{Sawa} T., {Fujimoto} M., 2005, PASJ, 57, 429

\bibitem[{{Sharma} {et~al.}(2016){Sharma}, {Theuns}, {Frenk}, {Bower}, {Crain},
  {Schaller}, \& {Schaye}}]{sharma16}
{Sharma} M., {Theuns} T., {Frenk} C., {Bower} R., {Crain} R., {Schaller} M.,
  {Schaye} J., 2016, \mnras, 458, L94

\bibitem[{{Shaya} \& {Tully}(2013)}]{shaya13}
{Shaya} E.~J., {Tully} R.~B., 2013, \mnras, 436, 2096

\bibitem[{{Slater} \& {Bell}(2013)}]{slater13}
{Slater} C.~T., {Bell} E.~F., 2013, \apj, 773, 17

\bibitem[{{Smith} {et~al.}(2016){Smith}, {Duc}, {Bournaud}, \& {Yi}}]{smith15}
{Smith} R., {Duc} P.~A., {Bournaud} F., {Yi} S.~K., 2016, \apj, 818, 11

\bibitem[{{Spergel} {et~al.}(2007){Spergel}, {Bean}, {Dor{\'e}}, {Nolta},
  {Bennett}, {Dunkley}, {Hinshaw}, \& {Wright}}]{spergel07}
{Spergel} D.~N., {Bean} R., {Dor{\'e}} O., {Nolta} M.~R., {Bennett} C.~L.,
  {Dunkley} J., {Hinshaw} G., {Wright} E.~L., 2007, \apjs, 170, 377

\bibitem[{{Springel}(2005)}]{springel05}
{Springel} V., 2005, \mnras, 364, 1105

\bibitem[{{Springel} {et~al.}(2008){Springel}, {Wang}, {Vogelsberger},
  {Ludlow}, {Jenkins}, {Helmi}, {Navarro}, {Frenk}, \& {White}}]{springel08}
{Springel} V., {Wang} J., {Vogelsberger} M., {Ludlow} A., {Jenkins} A., {Helmi}
  A., {Navarro} J.~F., {Frenk} C.~S., {White} S.~D.~M., 2008, \mnras, 391, 1685

\bibitem[{{Strigari} {et~al.}(2008){Strigari}, {Bullock}, {Kaplinghat},
  {Simon}, {Geha}, {Willman}, \& {Walker}}]{strigari08}
{Strigari} L.~E., {Bullock} J.~S., {Kaplinghat} M., {Simon} J.~D., {Geha} M.,
  {Willman} B., {Walker} M.~G., 2008, \nat, 454, 1096

\bibitem[{{Strigari} {et~al.}(2010){Strigari}, {Frenk}, \&
  {White}}]{strigari10}
{Strigari} L.~E., {Frenk} C.~S., {White} S.~D.~M., 2010, \mnras, 408, 2364

\bibitem[{{Tamm} {et~al.}(2012){Tamm}, {Tempel}, {Tenjes}, {Tihhonova}, \&
  {Tuvikene}}]{tamm12}
{Tamm} A., {Tempel} E., {Tenjes} P., {Tihhonova} O., {Tuvikene} T., 2012, \aap,
  546, A4

\bibitem[{{Tempel} {et~al.}(2015){Tempel}, {Guo}, {Kipper}, \&
  {Libeskind}}]{tempel15}
{Tempel} E., {Guo} Q., {Kipper} R., {Libeskind} N.~I., 2015, \mnras, 450, 2727

\bibitem[{{Teyssier} {et~al.}(2012){Teyssier}, {Johnston}, \&
  {Kuhlen}}]{teyssier12}
{Teyssier} M., {Johnston} K.~V., {Kuhlen} M., 2012, \mnras, 426, 1808

\bibitem[{{Tully} {et~al.}(2015){Tully}, {Libeskind}, {Karachentsev},
  {Karachentseva}, {Rizzi}, \& {Shaya}}]{tully15}
{Tully} R.~B., {Libeskind} N.~I., {Karachentsev} I.~D., {Karachentseva} V.~E.,
  {Rizzi} L., {Shaya} E.~J., 2015, \apjl, 802, L25

\bibitem[{{Tully} {et~al.}(2006){Tully}, {Rizzi}, {Dolphin}, {Karachentsev},
  {Karachentseva}, {Makarov}, {Makarova}, {Sakai}, \& {Shaya}}]{tully06}
{Tully} R.~B., {Rizzi} L., {Dolphin} A.~E., {Karachentsev} I.~D.,
  {Karachentseva} V.~E., {Makarov} D.~I., {Makarova} L., {Sakai} S., {Shaya}
  E.~J., 2006, \aj, 132, 729

\bibitem[{{Walker} \& {Loeb}(2014)}]{walker14}
{Walker} M.~G., {Loeb} A., 2014, Contemporary Physics, 55, 198

\bibitem[{{Walker} {et~al.}(2007){Walker}, {Mateo}, {Olszewski}, {Gnedin},
  {Wang}, {Sen}, \& {Woodroofe}}]{walker07}
{Walker} M.~G., {Mateo} M., {Olszewski} E.~W., {Gnedin} O.~Y., {Wang} X., {Sen}
  B., {Woodroofe} M., 2007, \apjl, 667, L53

\bibitem[{{Wang} {et~al.}(2013){Wang}, {Frenk}, \& {Cooper}}]{wang13}
{Wang} J., {Frenk} C.~S., {Cooper} A.~P., 2013, \mnras, 429, 1502

\bibitem[{{Wetzel} {et~al.}(2015){Wetzel}, {Deason}, \&
  {Garrison-Kimmel}}]{wetzel15}
{Wetzel} A.~R., {Deason} A.~J., {Garrison-Kimmel} S., 2015, \apj, 807, 49

\bibitem[{{Wetzel} {et~al.}(2013){Wetzel}, {Tinker}, {Conroy}, \& {van den
  Bosch}}]{wetzel13}
{Wetzel} A.~R., {Tinker} J.~L., {Conroy} C., {van den Bosch} F.~C., 2013,
  \mnras, 432, 336

\bibitem[{{Wheeler} {et~al.}(2015){Wheeler}, {O{\~n}orbe}, {Bullock},
  {Boylan-Kolchin}, {Elbert}, {Garrison-Kimmel}, {Hopkins}, \& {Kere{\v
  s}}}]{wheeler15}
{Wheeler} C., {O{\~n}orbe} J., {Bullock} J.~S., {Boylan-Kolchin} M., {Elbert}
  O.~D., {Garrison-Kimmel} S., {Hopkins} P.~F., {Kere{\v s}} D., 2015, \mnras,
  453, 1305

\bibitem[{{White} \& {Frenk}(1991)}]{white91}
{White} S.~D.~M., {Frenk} C.~S., 1991, \apj, 379, 52

\bibitem[{{White} \& {Rees}(1978)}]{white78}
{White} S.~D.~M., {Rees} M.~J., 1978, \mnras, 183, 341

\bibitem[{{Widrow} \& {Dubinski}(2005)}]{widrow05}
{Widrow} L.~M., {Dubinski} J., 2005, \apj, 631, 838

\bibitem[{{Yozin} \& {Bekki}(2015)}]{yozin15}
{Yozin} C., {Bekki} K., 2015, \mnras, 453, 2302

\bibitem[{{Zentner} {et~al.}(2005){Zentner}, {Kravtsov}, {Gnedin}, \&
  {Klypin}}]{zentner05}
{Zentner} A.~R., {Kravtsov} A.~V., {Gnedin} O.~Y., {Klypin} A.~A., 2005, \apj,
  629, 219

\end{thebibliography}
\end{document}